\documentclass[aip,pop,reprint]{revtex4-2}%
\usepackage{amsmath}
\usepackage{amsfonts}
\usepackage{amssymb}
\usepackage{graphicx}
\usepackage{verbatim}
\usepackage{dcolumn}
\usepackage{bm}%

\begin{document}

\title{Unified fluid-model theory of $\mathbf{E}\times\mathbf{B}$ instabilities
in low-ionized collisional plasmas with arbitrarily magnetized multi-species ions}
\author{Y. S. Dimant}
\author{M. M. Oppenheim}
\author{S. Evans}
\affiliation{Boston University, 725 Commonwealth Ave., Boston, MA 02215, USA}
\author{J. Martinez-Sykora}
\affiliation{Lockheed Martin Solar \& Astrophysics Laboratory, 3251 Hanover St, Palo Alto, CA 94304, USA}
\date{\today}

\begin{abstract}
This paper develops a unified linear theory of local cross-field plasma
instabilities, such as the Farley-Buneman, electron thermal, and ion thermal
instabilities, in collisional plasmas with fully or partially unmagnetized
multi-species ions. Collisional plasma instabilities in low-ionized, highly
dissipative, weakly magnetized plasmas play an important role in the lower
Earth's ionosphere and may be of importance in other planet ionospheres, star
atmospheres, cometary tails, molecular clouds, accretion disks, etc. In the
solar chromosphere, macroscopic effects of collisional plasma instabilities
may contribute into significant heating --- an effect originally suggested
from spectroscopic observations and relevant modeling. Based on a simplified
5-moment multi-fluid model, the theoretical analysis produces the general
linear dispersion relation for the combined Thermal-Farley-Buneman Instability
(TFBI). Important limiting cases are analyzed in detail. The analysis
demonstrates acceptable applicability of this model for the processes under
study. Fluid-model simulations usually require much less computer resources
than do more accurate kinetic simulations, so that the apparent success of
this approach to the linear theory of collisional plasma instabilities makes
it possible to investigate the TFBI (along with its possible macroscopic
effects) using global fluid codes originally developed for large-scale
modeling of the solar and planetary atmospheres.
\end{abstract}

\maketitle

\section{INTRODUCTION\label{Introduction}}

This paper develops a unified linear theory of local cross-field plasma
instabilities, such as the Farley-Buneman instability (FBI)
\cite{Farley:Plasma63,Buneman:Excitation63}, electron thermal instability
(ETI) \cite{Dimant:Kinetic95b,Dimant:Physical97,Robinson:Effects:98,StMaurice:Role00},
and ion thermal instability (ITI)
\cite{Kagan:Thermal00,Dimant:Ion04}. These instabilities may occur in
low-ionized and highly dissipative plasmas embedded in crossed electric and
magnetic fields. Such conditions are typical for the lower (E-region) Earth's
ionosphere, solar chromosphere, other planetary ionospheres, and they could exist in
such low-ionized gaseous objects as cometary tails, molecular clouds,
accretion disks, etc. The above local instabilities, along with the nonlocal
gradient drift instability (GDI)
\cite{Hoh:Instability63,Maeda:Theoretical63,Simon:Instability63}, generate waves of
acoustic-like plasma density perturbations coupled with turbulent
electrostatic fields.

All these instabilities have been mostly studied with respect to the E-region
ionosphere, but the emphasis of this paper is on the solar chromosphere. The
chromosphere is a relatively cool interface between the warmer photosphere and
very hot corona. Any energy transferred from the surface of the Sun to the
corona necessarily goes though the chromosphere. Therefore, it is crucial to
understand this region and properly model its behavior. The solar chromosphere
is a highly complex and dynamic region where microphysics may play a
significant role. Recently, large improvements in observations and modeling
have been made. Radiative MHD models capture a large variety of chromospheric
dynamics, such as magneto-acoustic shocks \cite{Carlsson:NonLTE92,Wedemeyer:Numerical2004}, spicules
\cite{Hansteen:Dynamic06,Martinez-Sykora:Generation17}, and flux emergence or
local dynamos \cite{Rempel:Numerical14}.

However, when comparing chromospheric observable profiles, such as MgII from
IRIS observations \cite{De-Pontieu:Interface14} and CaII from ground-based
observatories, with synthesis from the above models, the synthetic profiles
typically turn out to be narrower than the profiles deduced from observations
\cite{Carlsson:New19}. This discrepancy could have come from the lack of
turbulence in models, but the additional OI lines indicate that this is
insufficient \cite{Carlsson:What15}. Another possible scenario to explain the
discrepancy is mass load or heating. Comparison between IRIS and ALMA
observations with radiative MHD single-fluid models, which included
ion-neutral interaction effects and non-equilibrium ionization, suggests that
spicules in the models are still up to a few thousand degrees lower
\cite{Chintzoglou:ALMA21}.

Fontenla et al. \cite{Fontenla:Chromospheric05,Fontenla:Chromospheric08} proposed a
new heating mechanism that has not been included in the previous models. This
heating mechanism involves plasma turbulence and is based on the analogy
between the solar chromosphere and the lower Earth's ionosphere. In the
latter, collisional cross-field instabilities leading to palpable plasma
turbulence have been studied extensively using radar and rocket observations,
analytic theory, and supercomputer simulations. These instability-driven
turbulence produces an important macroscopic effect of strong anomalous
electron heating detected by radars
\cite{Foster:Simultaneous00,Bahcivan:Plasma07}. This effect has been explained
using analytic models and kinetic simulations \cite{Dimant:Model03,Milikh:Model03,Oppenheim:Kinetic13}.
Fontenla et al. \cite{Fontenla:Chromospheric05,Fontenla:Chromospheric08} suggested that the chromosphere
may include similar heating processes. These and other analyses
\cite{Gogoberidze:Farley09,Madsen:Multi14,Gogoberidze:Electrostatic14,Fletcher:Effects18,Oppenheim:Newly20}
suggested that the collisional cross-field plasma
instabilities can really be developed under the chromosphere conditions, so
that the proposed heating mechanism is plausible. The accurate theory of the
relevant plasma instabilities should help explain how, and by how much, this
mechanism could contribute to the chromospheric heating. The linear theory of
these instabilities, developed in this paper, is a necessary step in that direction.

In a number of important aspects, the physical conditions of
the solar chromosphere are similar to those of the E-region ionosphere. Among
the common features are the low ionization and prevalence of plasma-neutral
collisions in such a way that electrons are still magnetized, while ions are
partially or fully unmagnetized due to their frequent collisions with neutral
particles (by magnetized $s$-species plasma\ particles we mean particles whose
gyrofrequency $\Omega_{s}$ is much larger than the ion-neutral mean collision
frequency $\nu_{sn}$, while by unmagnetized or partially unmagnetized
$s$-species we mean the opposite case of $\Omega_{s}\lesssim\nu_{sn}$). The
energy source for the instabilities is the DC electric field $\vec{E}_{0}$
perpendicular to the magnetic field $\vec{B}_{0}$, in the frame of reference
attached to the neutral-particle flow. If $\vec{E}_{0}$ is strong enough then
the above magnetization conditions lead to cross-field instabilities. In the
Earth's ionosphere, strong electric fields are either generated by a
neutral-atmosphere dynamo (in the equatorial E region) or are mapped from the
magnetosphere down to the high-latitude E region during geomagnetic storms and
other intense events. In the core of the solar chromosphere, where the ideal
MHD\ conditions do not apply, high-speed neutral flows decoupled from the
magnetic field and crossing the latter under a significant angle may exist
\cite{Leake:Ionized14,Martinez:Role15,Ballester:Partially:18,Soler:Theory22}. This translates to the occurrence of strong electric
fields in the neutral-flow frame of reference.

On the other hand, the E-region ionosphere and solar chromosphere have
noticeable distinctions, such as the differences in the ion and neutral
compositions. In the E-region ionosphere, the two major ion species have
fairly close molecular masses and collision characteristics, so that to a
reasonable accuracy they can be treated as one unified ion species. A totally
different situation takes place in the solar chromosphere. The ion composition
there may be quite diverse. While the neutral part is mostly H (for
simplicity, we ignore here a small contribution of neutral He
\cite{Asplund:Chemical09,Wargnier:Detailed22}), the dominant ions are not
necessarily protons, H$^{+}$. The ion composition is often dominated by
ionized metal and other heavy impurities (C$^{+}$, Mg$^{+}$, Si$^{+}$,
Fe$^{+}$, etc.) because the ionization potentials of the corresponding neutral
atoms are usually significantly lower than that of H. As a result, the
magnetization of various ions may differ dramatically. At a given location,
some ion species can be magnetized, while other species are fully or partially
unmagnetized \cite{Fletcher:Effects18,Oppenheim:Newly20}. The multi-species
ion composition with different magnetization characteristics modifies the
conditions of the plasma instability development and complicates their analysis.

Additionally, unlike the lower Earth's ionosphere where the dominant ions
(O$_{2}^{+}$, NO$^{+}$) and neutrals (N$_{2}$, O$_{2}$) are molecules, the
solar chromosphere consists mostly of atoms. In the E-region ionosphere,
within the characteristic range of the characteristic low energies
$\lesssim0.3~$eV, electron collisions with neutral molecules, due to the
excitation of rotational and vibrational molecular levels, lead to mostly
inelastic energy losses. In the solar chromosphere, the electron collisional
energy losses are supposed to be mostly elastic since, within the relevant
energy range of $\lesssim1$~eV, the excitation of the atomic electron levels
is almost negligible. Using the same arguments, we can safely presume that the
contribution of the non-equilibrium ionization
\cite{Leenaarts:Non-equilibrium07,Golding:Non-equilibrium16,Martinez-Sykora:Ion-neutral20}
is also relatively small. This has serious implications for the electron
temperature balance and instability generation, as we discuss in
Sec.\texttt{~}\ref{Background flows and ohmic heating}.

Finally, the chromospheric magnetic fields are much larger than the
geomagnetic field, as well as the chromospheric values of the plasma and
neutral temperatures are significantly higher than those in the Earth's
ionosphere. However, these and similar parameters are scalable, so that this
quantitative distinction is not a real problem for the theory.

To simulate the above instabilities in both the initial (linear) and later
(nonlinear) stages, one can use fluid-model, kinetic, or hybrid approaches.
Most accurate is the kinetic approach, especially that based on
particle-in-cell (PIC) codes
\cite{Janhunen:Perpendicular94,Oppenheim:Ion04,Oppenheim:Large08,Oppenheim:Kinetic13,Oppenheim:Newly20}.
Such codes usually include all relevant physics, but they typically require substantial computer resources.
At present time, the PIC codes can simulate only restricted local plasma
volumes during a limited time duration, and those scales are still orders of
magnitude smaller than the chromospheric features observed with the current
resolution. At the same time, simulations based on simplified fluid-model
equations are usually much less restrictive and can efficiently model even
global plasma environments, such as, e.g., supergranular scales of the lower
solar atmosphere and even entire planetary ionospheres.

Typical wave periods and wavelengths of turbulence generated as a result of
collisional plasma instabilities are usually larger than the inverse collision
frequencies and mean free paths, respectively. Plasma processes with such
temporal and spatial scaling are usually reasonably well described by
fluid-model equations, though particle kinetics can sometimes be of paramount
importance. Indeed, the growth rate $\gamma$ of the pure FBI increases with
the wavenumber $k$ as $\gamma\propto k^{2}$ until the wavelength becomes
comparable to the inverse ion-neutral collisional mean free path. For shorter
wavelengths (i.e., larger $k$), the kinetic effect of ion Landau damping
overcomes the $k^{2}$ increase of $\gamma$ and sharply turns it down to
negative values, thus providing total stabilization of the short-wavelength
waves, see, e.g.,~Ref.~\onlinecite{Ossakow:Parallel75}. As a result of this
competition, the maximum instability growth rate is typically reached at an
intermediate spectral range between the highly and weakly collisional bands
which are determined by the low and high ratios of the wavelength to the
collisional mean free path, respectively.

The theoretical approach of this paper is based on a simplified 5-moment
multi-fluid set of equations. This model includes automatically all relevant
mechanisms of the instability driving and dissipation, except the Landau
damping and a number of other, mostly inconsequential, factors. For the
ionospheric conditions, in the framework of the two-fluid model (electrons and
single-species ions) such fluid-model analysis has been performed recently in
a series of papers by Makarevich, see Ref. \onlinecite{Makarevich:Toward21} and
references therein. Makarevich studied the linear theory of the FBI, GDI, and
ITI (but not the ETI)\ for arbitrary wavelengths, regardless of the fact that
the short-wavelength band is beyond the applicability of the fluid model.

In this paper, bearing in mind mostly the conditions of the solar
chromosphere, we analyze the general case of multi-species ions with an
arbitrary degree of the ion-species magnetization. Furthermore, in the
E-region research it is usually implied that the FBI is the dominant and the
most energetically efficient instability, solely responsible for the anomalous
electron heating. The main reason why we also included in our present theory
the thermal instabilities is as follows. Our recent PIC simulations of plasma
instabilities under the chromospheric conditions revealed, to our surprise,
that the ETI is very important and can even dominate in some regions of the
solar chromosphere \cite{Oppenheim:Newly20}. As far as the ITI is concerned,
our previous research has demonstrated that the ion thermal driving usually
accompanies the FBI \cite{Dimant:Ion04,Oppenheim:Ion04} and hence needs to also be
included for consistency.

Our theoretical analysis produces the general linear fluid-model dispersion
relation for the combined Thermal-Farley-Buneman Instability (TFBI)\ that
includes all relevant driving mechanisms (except the nonlocal GDI). Our major
thrust is on the long-wavelength limit in which all collisional plasma
instabilities reach their minimum threshold. This limit is of special
importance because if at a given location the driving electric field is below
the minimum threshold value then this location is linearly stable for any
waves. Although the fluid model is rigorously valid only in the
long-wavelength limit, in some cases it is possible, following Makarevich
\cite{Makarevich:Toward21}, to extend the fluid-model treatment to all
wavelengths. In Appendix~A, we demonstrate that in spite of the total
absence of Landau damping the simplified 5-moment model provides stabilization
of sufficiently short-wavelength waves (though the fluid-model results may be
inaccurate there). This fact allows one to safely use fully fluid-model
equations to simulate all instabilities without fearing that the corresponding
code might \textquotedblleft explode\textquotedblright\ within the
short-wavelength band because of the absence of Landau damping.

This analytical theory provides predictions of the instability generation
threshold conditions and growth rates, depending on the specific local
parameters of the plasma media. Also, we demonstrate that the fluid-model
approach for simulating the TFBI is reasonably well justified, even without
including the important kinetic effect of Landau damping. This guarantees that
the global fluid codes developed for the large-scale modeling can be applied
to the simulation of the small-scale cross-field plasma instabilities as well.
The results of this analytic theory can serve as a guide for such simulations
and help analyze their results.

This paper is organized as follows. In Sec.~\ref{Initial equations}, we
introduce the initial equations. In
Sec.~\ref{Background flows and ohmic heating}, we describe the background
plasma affected by the given imposed electric field in the neutral frame of
reference. More specifically, we describe the mean particle flows
(Sec.~\ref{Mean particle flows}) and ohmic heating (Sec.~\ref{Ohmic heating}).
The knowledge of the accurate values of these background parameters is crucial
for the instability linear theory. In Sec.~\ref{Linear wave perturbations: General dispersion relation}, we
consider this linear theory and derive the general multi-fluid dispersion
relation for the TFBI. In Sec.~\ref{Long-wavelength limit}, which is central
to this paper, we study the most important limit of long-wavelength waves,
which is responsible for the minimum instability threshold. In this limit, to
the zeroth-order approximation, we derive the wave phase-velocity relation,
which is common for all instabilities (Sec.~\ref{Wave phase velocity relation}).
To the first-order approximation, in
Sec.~\ref{Instability growth rates: physical mechanisms}, we derive the
instability driving/damping rates, where separate terms describe the driving
mechanisms for each distinct collisional instability and for the total losses.
Section~\ref{Threshold electric field} discusses the most important
quantitative result of the linear theory of instabilities, i.e., the
instability threshold.
Section~\ref{Instability dispersion relation for arbitrary wavelengths}
discusses the general dispersion relation for arbitrary wanelengths.
Section~\ref{SUMMARY AND CONCLUSIONS} summarizes the paper results.
Appendix~A\ discusses the short-wavelength limit of the general dispersion
relation. The analysis of the short-wavelength limit guarantees that the
employed fluid model, even without Landau damping, can be safely used for
instabillity modeling at all wavelengths with no need for additional damping
mechanisms to stabilize the wave behavior at short wavelengths. Appendix~B
lists major notations used in the paper.

\section{INITIAL\ EQUATIONS\label{Initial equations}}

In low-ionized plasmas, the dominant neutral component is usually weakly
disturbed by the plasma turbulence, so that within small and short-duration
characteristic spatiotemporal scales of instabilities we assume the neutral
atmosphere to be spatially uniform and stationary. For simplicity, we will
consider the constant neutral background composed of a single-species gas.

The simplest, 5-moment, multi-fluid model includes the continuity, momentum,
and energy-balance equations. In the frame of reference moving with the local
neutral flow (assumed to be spatially uniform and stationary, as stated
above), for each plasma species fluid marked by the subscript $s$, these
equations can be written as
\begin{subequations}
\label{my_fluid_equations}%
\begin{align}
&  \frac{\partial n_{s}}{\partial t}+\nabla\cdot(n_{s}\vec{V}_{s}%
)=0,\label{my_fluid_equations_cont}\\
&  m_{s}\frac{D_{s}\vec{V}_{s}}{Dt}=q_{s}(\vec{E}+\vec{V}_{s}\times\vec
{B})-\frac{\nabla(n_{s}T_{s})}{n_{s}}-m_{s}\nu_{sn}\vec{V}_{s}%
,\label{my_fluid_equations_mom}\\
&  n_{s}^{2/3}\ \frac{D_{s}}{Dt}\left(  \frac{T_{s}}{n_{s}^{2/3}}\right)
=\frac{2}{3}\ M_{sn}\nu_{sn}\vec{V}_{s}^{2}-\delta_{sn}\nu_{sn}(T_{s}-T_{n}),
\label{my_fluid_equations_temperat}%
\end{align}
where $D_{s}/Dt=\partial/\partial t+\vec{V}_{s}\cdot\nabla$ is the substantial
derivative along the $s$-flow; $n_{s}$, $m_{s}$, $q_{s}$, $T_{s}$, and
$\vec{V}_{s}$ are the $s$-species particle number density, mass, electric
charge, temperature (in energy units), and mean fluid velocity, respectively;
$\nu_{sn}\ $is the mean momentum transfer frequency of an $s$-particle
collision with a neutral ($n$) particle, $M_{sn}=m_{s}m_{n}/(m_{s}+m_{n})$ is
the corresponding effective mass, and $\delta_{sn}$ is the mean collisional
energy-loss fraction (the notation $\delta_{sn}$ should not be confused with the Kronecker delta
function). For purely elastic collisions, we have $\delta_{sn}=2m_{s}%
/(m_{s}+m_{n})$. In the lower Earth's ionosphere, however, the energy losses
are dominated by inelastic electron-neutral (e-n) collisions determined mostly
by low-energy molecular rotational and vibrational excitations, so that
$\delta_{en}$ can be electron-velocity dependent and significantly larger than
the elastic value (though still $\delta_{en}\ll1$). In the solar chromosphere,
we presume $\delta_{sn}$ to be close to its elastic value. Further, $\vec{E}$
and $\vec{B}$ are the total electrostatic field\ and an imposed external
magnetic field\ respectively (both in the neutral-gas frame of reference).
Implying sufficiently small-scale and short-period wave perturbations, we
assume the large-scale local background magnetic field $\vec{B}(\vec{r},t)$ to
be spatially uniform, stationary, and sufficiently strong, so that its wave
perturbations caused by turbulent electric currents and non-electrostatic
electric fields can be neglected, $\vec{B}\approx\vec{B}_{0}$. For electrons,
the particle charge is $q_{e}=-e$, where $e$ is the elementary charge. In the
lower ionosphere, the ions are singly charged, $q_{i}=e$. For the solar
chromosphere, however, we cannot exclude the possibility of multiply charged
ions, so that we will keep the general average charge $q_{j}$ for each ion
species $j$. Within a given ion species there may be the whole spectrum of
discrete particle charges, so that, in principle, the average charge ratio
$q_{j}/e$ may have a non-integer value $\geq1$.

The simplified fluid-model set of Eq.~(\ref{my_fluid_equations}) implies that
the $s$-particle velocity distribution, along with its wave perturbations, are
reasonably close to Maxwellian. This set of equations includes all essential
factors crucial for the instability generation and damping, such as the
particle inertia in the left-hand side (LHS) of
Eq.~(\ref{my_fluid_equations_mom}), Lorentz force, pressure gradients, and
collisional friction ($-m_{s}\nu_{sn}\vec{V}_{s}$) in the right-hand side
(RHS) of Eq.~(\ref{my_fluid_equations_mom}), the heat advection and adiabatic
heating/cooling in the LHS of Eq.~(\ref{my_fluid_equations_temperat}), as well
as even more important local collisional heating and cooling in the RHS of
Eq.~(\ref{my_fluid_equations_temperat}). The somewhat unconventional form of
energy-balance Eq.~(\ref{my_fluid_equations_temperat}) with its LHS
proportional to the substantial derivative of the specific enthropy
($T_{s}/n_{s}^{2/3}$) is more convenient for our purposes. In particular, this
form explicitly shows that in the absence of the collisional heating and
cooling -- the first and second terms of the right-hand side (RHS)
respectively -- the particle temperature obeys the adiabatic temperature
regime, $T_{s}\propto n_{s}^{2/3}$.

Equation~(\ref{my_fluid_equations}) neglects a number of known factors that
are largely inconsequential for the processes under study, largely due to the
aforementioned constraints on the typical turbulence spatial and temporal
scales. Among the major neglected factors are: Coulomb collisions between the
charged particles, slow processes of ionization and plasma annihilation
(recombination), pressure anisotropy (viscosity), higher moments of the
particle velocity distributions, the gravity force, and heat conductivity.

In the equatorial and high-latitude E-region ionospheres, the electrojet
instabilities are driven by an imposed significant DC electric field $\vec
{E}_{0}$. Its scales of spatial and temporal variation are usually much larger
than the characteristic wave scales, so that one may treat $\vec{E}_{0}$ as
spatially uniform and constant. In the solar chromosphere, neutral flows that
originate from below the chromosphere may decouple from the magnetic field and
cross the magnetic field lines. In a local frame of the neutral flow moving
with the neutral mass velocity $\vec{V}_{n}$ across a given magnetic field,
$\vec{B}_{0}$, we have an external large-scale DC electric field $\vec{E}%
_{0}=-\vec{V}_{n}\times\vec{B}_{0}$. Then the total electrostatic field is
$\vec{E}=\vec{E}_{0}-\nabla\Phi$, where $\Phi$ is the electrostatic potential
produced by plasma turbulence. Poisson's equation for $\Phi(r,t)$,
\end{subequations}
\begin{equation}
\nabla^{2}\Phi=\frac{1}{\epsilon_{0}}\left(  en_{e}-\sum_{j=1}^{p}q_{j}%
n_{j}\right)  , \label{Poisson_general}%
\end{equation}
closes the electrostatic description of plasma dynamics (here the integer $p$
is the total number of the ion species; $\epsilon_{0}$ is the permittivity of
free space). Typical turbulent wavelengths are much larger than the Debye
lengths. This usually allows one to employ the quasi-neutrality relation,
$en_{e}=\sum_{j=1}^{p}q_{j}n_{j}$, which eliminates the need for Poisson's
equation and simplifies the treatment. Bearing in mind, however, that even
small deviations from the quasi-neutrality in plasma waves may sometimes be of
importance (as we discuss below), for the linear waves generated by the
instabilities we will use Eq.~(\ref{Poisson_general}). For the large-scale
background plasma density $n_{s}=n_{s0}$, we will assume the full local charge
neutrality,
\begin{equation}
en_{e0}=\sum_{j=1}^{p}q_{j}n_{j0}. \label{local_quasineutrality}%
\end{equation}

\section{BACKGROUND FLOWS AND MEAN OHMIC HEATING\label{Background flows and ohmic heating}%
}

The driving force of all collisional plasma instabilities is the external
DC\ electric field, $\vec{E}_{0}\perp\vec{B}_{0}$, that must exist in the
frame of reference attached to the neutral atmosphere. The collisional plasma
response to this driving field is twofold: the external field creates distinct
electron and ion particle flows (leading to an anisotropic electric current)
and it also heats the plasma through the friction caused by collisions of the
plasma with the neutral particles. On the one hand, the stronger is the field
$\vec{E}_{0}$ the faster are the particle flows and the better should be the
conditions for the instability excitation. On the other hand, a stronger field
$\vec{E}_{0}$ results in larger mean ohmic heating of the plasma. The elevated
electron and ion temperatures increase the plasma diffusion within the waves
and, through the increased instability threshold, make the heated plasma more
resistive to the instability excitation. If, nonetheless, the driving field
magnitude, $E_{0}=|\vec{E}_{0}|$, exceeds the increased instability threshold,
$E_{\mathrm{Thr}}$, then the linear instability will develop, but saturated
plasma turbulence will be less intense than it might be without such
macroscopic heating. In the non-linear stage, the turbulent electric field
additionally heats up plasma particles, affecting the saturated level of
developed turbulence. In this paper, however, we deal only with the initial
linear stage of instabilities.

\subsection{Mean particle flows\label{Mean particle flows}}

Consider the undisturbed background plasma embedded in the external
macroscopic electric ($\vec{E}_{0}$) and magnetic ($\vec{B}_{0}$) fields. For
a given plasma species $s$ (electrons or $j$-species ions),
Eq.~(\ref{my_fluid_equations_mom}) yields the following mean fluid velocity:
\begin{equation}
\vec{V}_{s0}=\left.\!\!  \left(  \frac{q_{s}\vec{E}_{0}}{m_{s}\nu_{sn}}+\kappa
_{s}^{2}\vec{V}_{0}\right)\!\!  \right/ \!\!\! \left(1+\kappa_{s}^{2}\right)
=\frac{\kappa_{s}(\vec{E}_{0}+\kappa_{s}\vec{E}_{0}\times\hat{b})}%
{(1+\kappa_{s}^{2})B_{0}}. \label{V_s0}%
\end{equation}
Here%
\begin{equation}
\vec{V}_{0}\equiv\frac{\vec{E}_{0}\times\vec{B}_{0}}{B_{0}^{2}}=\frac{\vec
{E}_{0}\times\hat{b}}{B_{0}} \label{V_0_again}%
\end{equation}
is the $\vec{E}_{0}\times\vec{B}_{0}$ drift velocity, where $\hat{b}\equiv
\vec{B}_{0}/B_{0}$ is the unit vector in the direction of $\vec{B}_{0}$,
$\Omega_{s}=q_{s}B_{0}/m_{s}$ is the $s$-species gyrofrequency, and
\begin{equation}
\kappa_{s}=\frac{\Omega_{s}}{\nu_{sn}}=\frac{q_{s}B_{0}}{m_{s}\nu_{sn}}
\label{kappa_s_again}%
\end{equation}
is the corresponding magnetization parameter. In this paper, we mostly imply
strongly magnetized electrons, $\kappa_{e}^{2}\gg1$, while a
multi-species positive-ion population, $s=j$, may contain both unmagnetized or magnetized ions.
In other words, we allow the ion magnetization to be weak, $\kappa_{j}\ll1$, or
moderate, $\kappa_{j}\gtrsim1$, but not strong (not $\kappa_{j}\gg1$). Strongly magnetized ions
are of no interest for the collisional instabilities, since for $\kappa_{j}>1$
the FBI mechanism becomes stabilizing with the stabilization facor increasing
proportionally with $(\kappa_{j}^{2}-1)$, see Ref.~\onlinecite{Dimant:Ion04}.

For each ion species $j$, we introduce the difference between the
undisturbed electron and ion drift velocities, $\vec{U}_{j}\equiv\vec{V}%
_{e0}-\vec{V}_{j0}$. We will actively use this parameter in the following
sections. Strongly magnetized electrons move with almost the $\vec{E}%
_{0}\times\vec{B}_{0}$ drift velocity, $\vec{V}_{e0}\approx\vec{V}_{0}$, so
that Eq.~(\ref{V_s0}) yields
\begin{equation}
\vec{U}_{j}\approx\vec{V}_{0}-\vec{V}_{j0}=\frac{\vec{V}_{0}-\kappa_{j}\vec
{E}_{0}/B_{0}}{1+\kappa_{j}^{2}}=\frac{\vec{E}_{0}\times\hat{b}-\kappa_{j}%
\vec{E}_{0}}{(1+\kappa_{j}^{2})B_{0}}. \label{UU}%
\end{equation}
Comparing the expression for the ion drift velocity from Eq.~(\ref{V_s0})
($s=j$) with Eq.~(\ref{UU}), we easily find that $\vec{V}_{j0}$ and $\vec
{U}_{j}$ are mutually orthogonal and relate to each other as $\vec{V}%
_{j0}\times\hat{b}=\kappa_{j}\vec{U}_{j}$. Bearing in mind that to the same
accuracy $\vec{U}_{j}+\vec{V}_{j0}=\vec{V}_{0}$, we obtain that the absolute
values of $\vec{V}_{0}$, $\vec{V}_{j0}$, and $\vec{U}_{j}$ relate to each
other as%
\begin{equation}
V_{j0}=\kappa_{j}U_{j},\qquad U_{j}=\frac{V_{0}}{\sqrt{1+\kappa_{j}^{2}}}.
\label{VVU_abs_values}%
\end{equation}
Through the magnetization parameter $\kappa_{j}$, the above relations depend
on the ion-neutral collisional frequency, $\nu_{jn}$. In the general case,
$\nu_{jn}$ might be temperature-dependent and hence could be modified by the
ohmic heating. However, throughout this paper we assume
temperature-independent ion-neutral collision frequencies, as we discuss right below.

For two colliding particles -- a charged particle $s$ and a neutral particle
$n$ -- the approximation of the constant collision frequency, $\nu_{sn}%
=n_{n}\sigma_{sn}V_{sn}$ is called ``Maxwell molecule
collisions'' (MMC) approximation \cite{Schunk:Ionospheres09}
(here $n_{n}$ is the $n$-particle density, $V_{sn}$ is the relative speed of the two
colliding particles during their initial remote approach for a given
collision, and $\sigma_{sn}$ is the $V_{sn}$-dependent $s$-$n$
collisional cross-section). After averaging over the entire particle velocity distributions, this leads
to the temperature-independent mean collision frequency $\nu_{sn}$. For
plasma-neutral collisions, the MMC approximation is usually based on the
assumption that the collision cross-sections are mostly determined by the
charged-particle-induced polarization of the neutral collision partner (the
corresponding interaction potential is $\propto1/r_{\mathrm{int}}^{4}$, where
$r_{\mathrm{int}}$ is the inter-particle distance). This results in the $s$-$n$
collision cross-section $\sigma_{sn}\propto1/V_{sn}$, so that the kinetic
collision frequency $\nu_{sn}$ becomes velocity-independent. In the solar
chromosphere where neutral particles are predominantly hydrogen atoms, within
the low-energy range of $\lesssim1~$eV the MMC approximation should work
reasonably well for both $e$-$n$ and $i$-$n$ collisions, except
proton-hydrogen (H$^{+}$-H) collisions, which are strongly affected by the
charge exchange. However, even for the latter, the MMC approximation still works
reasonably well. For both H$^{+}$-H and $e$-H collisions, this can be
verified, e.g., from the $\sigma_{H^{+}n}$ and $\sigma_{en}$ data presented in
Ref.\ \onlinecite{Vranjes:Collisions13}, Fig.~1 and 4 (after smoothening in Fig.~1 the
curves over frequent quantum oscillations, see also
Ref.~\onlinecite{Wargnier:Detailed22}). Assuming plasma collisions with hydrogen
atoms to be elastic, we will employ in the chromosphere the MMC approximation
for all $j$-$n$ and $e$-$n$ collisions. In the E-region ionosphere, however,
the dominant neutral particles are molecules. Within the relevant low-energy
range $\lesssim0.3$~eV, collisional losses of electron energy are dominated
by inelastic excitation of rotational and vibrational molecular levels. As a result,
in the ionosphere, the MMC approximation does not work
for the $e$-$n$ collisions \cite{Gurevich:Nonlinear78}, but for the
ion-neutral collisions it generally works reasonably well \cite{Schunk:Ionospheres09}. In
this paper, bearing in mind mostly the chromospheric conditions with
predominantly elastic \emph{e-n} collisions, we will assume constant $\nu
_{sn}$ for all \emph{e-n} and \emph{i-n} collisions.

\subsection{Ohmic heating\label{Ohmic heating}}

Now we discuss the large-scale frictional heating of plasma particles in the
crossed $\vec{E}_{0}$ and $\vec{B}_{0}$ fields. For the background temperature
of charged particles, Eqs.~(\ref{my_fluid_equations_temperat})
and~(\ref{VVU_abs_values}), lead to
\begin{equation}
T_{s0}=T_{n}+\frac{2M_{sn}\kappa_{s}^{2}V_{0}^{2}}{3\delta_{sn}\left(
1+\kappa_{s}^{2}\right)  }\approx T_{n}+\frac{m_{n}\kappa_{s}^{2}V_{0}^{2}%
}{3\left(  1+\kappa_{s}^{2}\right)  }, \label{T_s0_again}%
\end{equation}
where the far right approximate expression applies only to purely elastic
collisions with $\delta_{sn}=\delta_{sn}^{\mathrm{elas}}=2m_{s}/(m_{s}+m_{n}%
)$. Equation~(\ref{T_s0_again}) describes the background ohmic caused by the
driving electric field $\vec{E}_{0}$.

For strongly magnetized electrons, $\kappa_{e}^{2}\gg1$, Eq.~(\ref{T_s0_again}%
) reduces to%
\begin{equation}
T_{e0}=T_{n}+\frac{2m_{e}V_{0}^{2}}{3\delta_{en}}\approx T_{n}+\frac
{m_{n}V_{0}^{2}}{3}, \label{T_e0}%
\end{equation}
where, as above, the far right expression applies only to elastic
electron-neutral collisions with $\delta_{en}=\delta_{en}^{\mathrm{elas}%
}\approx2m_{e}/m_{n}$.

Equation (\ref{T_e0})\ has a serious implication for the instability driving.
To drive a collisional instability, like the FBI, one needs to apply an
external DC electric field $\vec{E}_{0}\perp\vec{B}_{0}$. This field
amplitude, $E_{0}$, must exceed the minimum threshold value, $E_{\mathrm{Thr}%
}^{\min}$, assuming that instability driving overcomes the regular plasma
diffusion caused by the plasma pressure gradients within the generated waves.
For example, in a single-species ion (SSI)\ plasma ($j=i$), the minimum
FBI threshold field corresponds to the $\vec{E}_{0}\times\vec{B}_{0}$ speed
close to the isothermal ion acoustic speed, $C_{s}$,
\begin{equation}
V_{0}\approx C_{s}\equiv\left(  \frac{T_{e0}+T_{i0}}{m_{i}}\right)  ^{1/2}.
\label{V_0_approx}%
\end{equation}
According to Eqs.~(\ref{T_s0_again}) (for $s=i$) and (\ref{T_e0}), the driving
field heats both ions and electrons, increasing the instability threshold.
Under the optimum conditions for the FBI with essentially unmagnetized ions,
$\kappa_{i}^{2}\ll1$, the ion heating is usually moderate and not detrimental
for the instability excitation.

A totally different situation takes place for electrons. For the E-region
Earth's ionosphere with dominant molecular ions (NO$^{+}$, O$_{2}^{+}$) the
electron energy loss rate, $\delta_{en}$, is determined mostly by inelastic
losses caused by collisional excitation of low-energy rotational and
vibrational molecular\ levels. The corresponding inelastic
temperature-dependent parameter, $\delta_{en}=\delta_{en}^{\mathrm{inel}}$,
still remains small, $\delta_{en}^{\mathrm{inel}}\simeq(2$--$4)\times10^{-3}$, see
Ref. \onlinecite{Gurevich:Nonlinear78}, but two orders of magnitude larger than the
corresponding elastic value, $\delta_{en}^{\mathrm{elas}}\approx2m_{e}%
/m_{n}\simeq3.5\times10^{-5}$ (assuming the N$_{2}$, O$_{2}$-dominated Earth's
neutral atmosphere). The corresponding ohmic heating described by the middle
expression in Eq.~(\ref{T_e0}) with $\delta_{en}=\delta_{en}^{\mathrm{inel}}$
is noticeable, but still not detrimental for the FBI excitation. A drastically
different situation, however, should take place in the atomic gas atmosphere,
such as the solar chromosphere where the hydrogen (H) prevails in the neutral
atmosphere. Atoms have no rotational or vibrational losses, and for typical
chromospheric temperatures below 1~eV we expect no significant excitation of
the electronic levels. Indeed, excitation of the lowest excited atomic state
requires 10.2~eV, so that for $T_{e}=11,600~$K (corresponding to 1~eV), the
fraction of Maxwellian superthermal electrons that may provide such excitation
is $\sim\sqrt{10.2}\exp(-10.2)\simeq10^{-4}$. The fraction of electrons that
can ionize the neutral H atoms is even smaller, $\sim13.6\exp(-13.6)\simeq
2\times10^{-5}$. The fractions of the total energy losses corresponding to
these inelastic processes are roughly given by the same numbers. As a matter
of fact, relevant chromospheric temperatures are usually smaller,
$\lesssim0.5$~eV, so that the inelastic energy loss fractions are even
exponetially smaller than those estimated above. Comparing these small
fractions with the mean elastic energy loss fraction $\delta_{en}%
^{\mathrm{elas}}\approx2m_{e}/m_{\mathrm{H}}\simeq10^{-3}$, we see that
inelastic electron-energy losses, including those associated with the
non-equilibrium ionization
\cite{Leenaarts:Non-equilibrium07,Golding:Non-equilibrium16,Martinez-Sykora:Ion-neutral20}%
, can be neglected. Under these assumptions, the collisional energy loss
fraction $\delta_{en}$ should be reasonably close to its elastic value,
$\delta_{en}^{\mathrm{elas}}$. Then the corresponding ohmic heating is
determined by the far right expression in Eq.~(\ref{T_e0}). According to it,
the ratio of $E_{0}$ to the temperature-modified minimum FBI threshold,
$E_{\mathrm{Thr}}^{\min}$, is determined by%
\begin{equation}
\frac{E_{0}}{E_{\mathrm{Thr}}^{\min}}=\frac{V_{0}}{C_{s}}=\sqrt{\frac
{3m_{i}\left(  T_{e}-T_{n}\right)  }{m_{n}\left(  T_{e}+T_{i}\right)  }}.
\label{7}%
\end{equation}
If all ions were created by ionizing the dominant neutral gas atoms or
molecules, with no further chemical reactions, then we would have $m_{i}%
=m_{n}$. In such cases, regardless of how strong is the driving electric field
$\vec{E}_{0}$, the ratio $E_{0}/E_{\mathrm{Thr}}^{\min}$ could not exceed a
fairly modest value of $\sqrt{3}\approx1.73$ (corresponding to $T_{e}%
\rightarrow\infty$). In the lower ionosphere, even in spite of the slightly
different neutral and ion molecular compositions, the approximate equality,
$m_{i}\approx m_{n}$, holds. This means that if there were no rotational and
vibrational energy losses then ohmic heating by the driving field would be so
high that the FBI\ could only be excited within a narrow altitude range with
only a moderate increase of the driving field above the temperature-modified
threshold value. However, in the solar chromosphere, where the neutral
composition is mostly H, but small impurities with the low ionization
potential become ionized much easier than H, the much heavier metal ions can
become a significant, if not dominant, fraction of the ionized component. As
a result, the average ion mass $m_{i}$ may exceed $m_{n}$ by a noticeable
factor. This helps the ratio $E_{0}/E_{\mathrm{Thr}}^{\min}$ reach far larger
values than $\sqrt{3}$ and hence lead to more intense plasma turbulence.

This discussion is based on a simplified model that assumes just one kind of
instability (FBI), but the same basic idea applies to the more general and
complicated situation. The important point is that one has to
self-consistently account for possible modifications of the background plasma
caused by the driving field itself because some of these modifications can
improve or aggravate the instability driving conditions.

\section{LINEAR WAVE\ PERTURBATIONS\label{Linear wave perturbations: General dispersion relation}}

Now we start developing the linear theory of dissipative instabilities,
assuming the neutral-flow local frame of reference. The thrust of this section
is the derivation of the general dispersion relation using the 5-moment
multi-fluid model equations.

For all varying vector or scalar quantities, we will assume small harmonic
wave perturbations $\propto\exp[i(\vec{k}\cdot\vec{r}-\omega t)]$, where the
vector $\vec{k}$ is real, while the wave frequency, $\omega$, can be a complex
number: $\omega=\omega_{r}+i\gamma$ (with real $\omega_{r}$ and $\gamma$). In
this ansatz, the linear instability means positive $\gamma$ (the growth rate),
while a stable situation means negative $\gamma$ (the damping rate). In what
follows, we will denote small linear perturbations of any scalar or vector
quantity by adding $\delta$ to the corresponding variable notation, bearing in
mind that every perturbation, denoted like $\delta A$, represents just one
isolated harmonic wave with the complex amplitude.

For any isolated linearized harmonic wave perturbation, we have $\partial
/\partial t\rightarrow-i\omega$,\ $\nabla\rightarrow i\vec{k}$,\ and
$\partial/\partial t+\vec{V}_{s0}\cdot\nabla\rightarrow-i\omega_{Ds}$, where
\begin{equation}
\omega_{Ds}\equiv\omega-\vec{k}\cdot\vec{V}_{s0} \label{Omega_alpha}%
\end{equation}
is the Doppler-shifted wave frequency in the frame of reference moving with
the $s$-species mean flow, $\vec{V}_{s0}$. We will separate the wavevector
$\vec{k}$ to its parallel (to $\vec{B}_{0}=B_{0}\hat{b}$) and perpendicular
components, $\vec{k}=k_{\parallel}\hat{b}+\vec{k}_{\perp}$. In what follows,
we will assume field-aligned wave perturbations, $k_{\perp}\equiv|\vec
{k}_{\perp}|\gg|k_{\parallel}|$, so that $k_{\perp}\approx k\equiv|\vec{k}|$.
Non-field-aligned wave modes with $|k_{\parallel}|\sim k_{\perp}$ are usually
situated deeply within the linearly stable range and are of no interest for
the linear instability analysis. However, even the small parallel component
$k_{\parallel}$ should be included in the theory because it may be of
importance for the electron dynamics and heating, see Ref.\ \onlinecite{Dimant:Model03} and
references therein.

Temporarily introducing dimensionless variables,%
\begin{equation}
\eta_{s}\equiv\frac{\delta n_{s}}{n_{s0}},\qquad\phi\equiv\frac{e\delta\Phi
}{T_{e0}},\qquad\tau_{s}\equiv\frac{\delta T_{s}}{T_{s0}}, \label{Introducing}%
\end{equation}
and linearizing the $s$-particle number density, velocity, temperature, and
electrostatic potential against their background values (discussed in the
preceding section), from continuity Eq.~(\ref{my_fluid_equations_cont}), we
obtain%
\begin{equation}
\eta_{s}=\frac{\vec{k}\cdot\delta\vec{V}_{s}}{\omega_{Ds}}.
\label{eta_s_again}%
\end{equation}
Similarly, thermal Eq.~(\ref{my_fluid_equations_temperat}) yields%
\begin{equation}
-i\omega_{Ds}\left(  \tau_{s}-\frac{2}{3}\ \eta_{s}\right)  =\frac{4M_{sn}%
\nu_{sn}}{3T_{s0}}(\vec{V}_{s0}\cdot\delta\vec{V}_{s})-\delta_{sn}\nu_{sn}%
\tau_{s} \label{from_Temper}%
\end{equation}
Below we show that in the dimensionless variables (\ref{Introducing}) the
fluid velocity perturbation $\delta\vec{V}_{s}$ is proportional to the linear
combination $\left(  \alpha_{s}\phi+\eta_{s}+\tau_{s}\right)  $, where
\begin{equation}
\alpha_{s}\equiv\frac{T_{e0}q_{s}}{T_{s0}e}, \label{alpha_s}%
\end{equation}
so that $\delta\vec{V}_{s}=$ $\left(  \alpha_{s}\phi+\eta_{s}+\tau_{s}\right)
\vec{K}_{s}$, where the vector $\vec{K}_{s}$ will be determined later using
momentum Eq.~(\ref{my_fluid_equations_mom}).

Indeed, for each species we can separate in the RHS of
Eq.~(\ref{my_fluid_equations_mom}) the two velocity-independent forces, i.e.,
the electric field and the pressure-gradient forces. The remaining two
velocity-dependent forces, i.e., the magnetic component of the Lorentz force
and collisional friction, can be re-arranged to the LHS. The combined
linearized wave component of the velocity-independent forces is proportional
to $\left(  \alpha_{s}\phi+\eta_{s}+\tau_{s}\right)  \vec{k}$, while the
corresponding harmonic component $\propto\delta\vec{V}_{s}$ in the re-arranged
LHS determines the linear tensor response to that. Explicitly resolving this
linear response, we obtain $\delta\vec{V}_{s}=$ $\left(  \alpha_{s}\phi
+\eta_{s}+\tau_{s}\right)  \vec{K}_{s}$ and find the vector $\vec{K}_{s}$,
whose explicit expressions will be given below by Eqs.~(\ref{V_s||}) and
(\ref{V_sperp}).

In terms of still unspecified $\vec{K}_{s}$, Eqs.~(\ref{eta_s_again}) and
(\ref{from_Temper}) yield
\begin{subequations}
\label{1,2}%
\begin{align}
\eta_{s} &  =\left(  \alpha_{s}\phi+\eta_{s}+\tau_{s}\right)  A_{s}%
,\label{1}\\
\mu_{s}\tau_{s}-\frac{2}{3}\ \eta_{s} &  =\left(  \alpha_{s}\phi+\eta_{s}%
+\tau_{s}\right)  B_{s},\label{2}%
\end{align}
\end{subequations}
where%
\begin{widetext}
\begin{equation}
A_{s}\equiv\frac{\vec{k}\cdot\vec{K}_{s}}{\omega_{Ds}},\qquad B_{s}\equiv
i\ \frac{4M_{sn}\nu_{sn}(\vec{V}_{s0}\cdot\vec{K}_{s})}{3T_{s0}\omega_{Ds}%
},\qquad\mu_{s}\equiv1+\frac{i\delta_{sn}\nu_{sn}}{\omega_{Ds}}.\label{ABmu}%
\end{equation}
\end{widetext}
Solving Eq.~(\ref{1,2}) for $\tau_{s}$ and $\eta_{s}$ in terms of $\phi$, we
obtain
\begin{equation}
\tau_{s}=\frac{1}{\mu_{s}}\left(  \frac{2}{3}+\frac{B_{s}}{A_{s}}\right)
\eta_{s},\qquad\eta_{s}=\alpha_{s}N_{s}\phi,\label{tau_s_via_eta_s}%
\end{equation}
where%
\begin{equation}
N_{s}\equiv\left(  1-A_{s}-\frac{2A_{s}+3B_{s}}{3\mu_{s}}\right)  ^{-1}%
A_{s}.\label{phi=snova}%
\end{equation}
Then, linearizing Poisson's Eq.~(\ref{Poisson_general}) in these variables, we
obtain:%
\begin{equation}
\sum_{j=1}^{p}\rho_{j}\eta_{j}-\eta_{e}=k^{2}\lambda_{De}^{2}\phi,\qquad
\rho_{j}=\frac{q_{j}n_{j0}}{en_{e0}},\label{Debye_snova}%
\end{equation}
where $\lambda_{De}=[\epsilon_{0}T_{e0}/(e^{2}n_{e0})]^{1/2}$ is the
\textquotedblleft electron\textquotedblright\ Debye length. Using
Eq.~(\ref{phi=snova}), we express all $\eta_{s}$ in terms of $\phi$ and then
substitute the results to Eq.~(\ref{Debye_snova}). This gives us an interim
dispersion relation,
\begin{equation}
1+\sum_{j=1}^{p}\frac{\rho_{j}\alpha_{j}N_{j}}{N_{e}}=\frac{k^{2}\lambda
_{D}^{2}}{N_{e}},\label{disperga_snova}%
\end{equation}
in terms of the parameters $A_{s}$ and $B_{s}$ defined by Eq.~(\ref{ABmu}).

The ultimate dispersion relation requires explicit expressions for $A_{s}$ and
$B_{s}$. To determine these expressions, we have to find $\delta\vec{V}_{s}$
from momentum Eq.~(\ref{my_fluid_equations_mom}). Linearizing
Eq.~(\ref{my_fluid_equations_mom}), we obtain:%
\begin{equation}
\left(  1-i\ \frac{\omega_{Ds}}{\nu_{sn}}\right)  \delta\vec{V}_{s}-\kappa
_{s}(\delta\vec{V}_{s}\times\hat{b})=-i\ \frac{\vec{k}V_{Ts}^{2}}{\nu_{sn}%
}\left(  \alpha_{s}\phi+\eta_{s}+\tau_{s}\right)  , \label{V=}%
\end{equation}
where $V_{Ts}=(T_{s0}/m_{s})^{1/2}$ is the mean chaotic speed of the
$s$-particle velocity distribution. Then for the parallel components of
linearly related $\delta\vec{V}_{s}$ and $\vec{K}_{s}$, we obtain%
\begin{equation}
K_{s\parallel}=\frac{\delta\vec{V}_{s\parallel}}{\alpha_{s}\phi+\eta_{s}%
+\tau_{s}}=-i\ \frac{\vec{k}_{\parallel}V_{Ts}^{2}}{\nu_{sn}\left(
1-i\omega_{Ds}/\nu_{sn}\right)  }. \label{V_s||}%
\end{equation}
After applying a \textquotedblleft cross\textquotedblright-product $\times
\hat{b}$ to Eq.~(\ref{V=}) and then eliminating $\delta\vec{V}_{s}\times
\hat{b}$ from both equations, we obtain for the dominant perpendicular
components:%
\begin{widetext}
\begin{equation}
\vec{K}_{s\perp}=\frac{\delta\vec{V}_{s\perp}}{\alpha_{s}\phi+\eta_{s}%
+\tau_{s}}=-i\ \frac{V_{Ts}^{2}}{\nu_{sn}}\ \frac{\left(  1-i\omega_{Ds}%
/\nu_{sn}\right)  \vec{k}_{\perp}+\kappa_{s}(\vec{k}_{\perp}\times\hat{b}%
)}{\left(  1-i\omega_{Ds}/\nu_{sn}\right)  ^{2}+\kappa_{s}^{2}}.
\label{V_sperp}%
\end{equation}
From these expressions, we obtain now the explicit general expressions for
$A_{s}$ and $B_{s}$:
\begin{equation}
A_{s}=-i\ \frac{V_{Ts}^{2}}{\nu_{sn}\omega_{Ds}}\left[  \frac{\left(
1-i\omega_{Ds}/\nu_{sn}\right)  k_{\perp}^{2}}{\left(  1-i\omega_{Ds}/\nu
_{sn}\right)  ^{2}+\kappa_{s}^{2}}+\frac{k_{\parallel}^{2}}{1-i\omega_{Ds}%
/\nu_{sn}}\right]  , \label{A_s_opiat}%
\end{equation}%
\begin{equation}
B_{s}=\frac{4m_{n}}{3\omega_{Ds}\left(  m_{n}+m_{s}\right)  }\ \frac{\left(
1-i\omega_{Ds}/\nu_{sn}\right)  (\vec{k}_{\perp}\cdot\vec{V}_{s0})-\kappa
_{s}\vec{k}_{\perp}\cdot(\vec{V}_{s0}\times\hat{b})}{\left(  1-i\omega
_{Ds}/\nu_{sn}\right)  ^{2}+\kappa_{s}^{2}}, \label{B_s_gen}%
\end{equation}
\end{widetext}
valid for all plasma species $s$. Specifically for the strongly magnetized
electrons, $\kappa_{e}^{2}\gg1$, we obtain simpler expressions:
\begin{subequations}
\label{A_e,B_e}%
\begin{align}
A_{e}  &  \approx-i\ \frac{k_{\perp}^{2}V_{Te}^{2}\left[  \left(
1-i\omega_{De}/\nu_{e}\right)  ^{2}+\kappa_{e}^{2}k_{\parallel}^{2}/k_{\perp
}^{2}\right]  }{\nu_{en}\omega_{De}\kappa_{e}^{2}(1-i\omega_{De}/\nu_{en}%
)},\label{A_e}\\
B_{e}  &  \approx\frac{4k_{\perp}V_{0}}{3\omega_{De}\kappa_{e}^{2}}\left[
\left(  1-\frac{i\omega_{De}}{\nu_{e}}\right)  \cos\theta-\kappa_{e}\sin
\theta\right]  , \label{B_e}%
\end{align}
where $\theta$ is the angle from $\vec{V}_{0}$ to $\vec{k}$ (often called
the `flow' angle). Similarly, for $j$-species ions, we have
\end{subequations}
\begin{equation}
B_{j}=\frac{4\kappa_{j}kU_{j}m_{n}}{3\omega_{Dj}\left(  m_{n}+m_{j}\right)
}\ \frac{\left(  1-i\omega_{Dj}/\nu_{jn}\right)  \sin\chi_{j}-\kappa_{j}%
\cos\chi_{j}}{\left(  1-i\omega_{Dj}/\nu_{jn}\right)  ^{2}+\kappa_{j}^{2}},
\label{B_j_snova}%
\end{equation}
where the angle $\chi_{j}=\theta+\arctan\kappa_{j}$ is unambiguously
determined by relations:
\begin{align}
\cos\chi_{j}  &  =\frac{\vec{k}\cdot\vec{U}_{j}}{kU_{j}}=\frac{\cos
\theta-\kappa_{j}\sin\theta}{\sqrt{1+\kappa_{j}^{2}}},\nonumber\\
\sin\chi_{j}  &  =\frac{\vec{k}\cdot\vec{V}_{j}}{kV_{j}}=\frac{\sin
\theta+\kappa_{j}\cos\theta}{\sqrt{1+\kappa_{j}^{2}}}. \label{ion_angles}%
\end{align}
Recall that according to Eq.~(\ref{VVU_abs_values}) we can also express
$U_{j}$ in (\ref{B_j_snova}) in terms of $V_{0}=E_{0}/B_{0}$ as $U_{j}%
=V_{0}/(1+\kappa_{j}^{2})^{1/2}$. Using Eqs.~(\ref{A_s_opiat}%
)--(\ref{ion_angles}) and substituting all $A_{s}$, $B_{s}$ into
(\ref{disperga_snova}), we obtain the general dispersion relation for
$\omega(\vec{k})$. This general equation was published earlier
\cite{Oppenheim:Newly20} without the derivation and further theoretical
analysis. In the following section, assuming the limit of sufficiently
long-wavelength waves, we reduce this equation to a simpler form, more useful
for the physical analysis and simple estimates.

Equation~(\ref{disperga_snova}), where $\mu_s$, $N_s$, $A_s$, and $B_s$ are given by Eqs.~(\ref{ABmu}), (\ref{phi=snova}), (\ref{A_s_opiat}), and (\ref{B_s_gen}),
represents the general dispersion relation. We caution that in the short-wavelength range this expression is
physically deficient due to lack of crucial Landau damping. The major value of this equation, however, is that it
allows one to simulate instabilities for the entire wave spectrum using the cheaper fluid code, just ignoring a non-physical behavior at the
short-wavelength band. For many years researchers, including ourselves, were
afraid that a fluid code without Landau damping may blow-up at
short-wavelength waves. In Appendix\ \ref{APPENDIX A}, however, we demonstrate that there is
no need to be afraid of that. Below we present the long-wavelength limit
solution, which is not physically deficient because in this limit the missed
kinetic effect of Landau damping plays no role.

\section{LONG-WAVELENGTH LIMIT (LWL)\label{Long-wavelength limit}}

This section discusses the most important limiting case of the long-wavelength
limit (LWL). We define this limit as the $\omega$, $k$-band, in which $k^{-1}$
are much larger than both the collisional mean free paths and the Debye lengths,
$\lambda_{Ds}$, while the wave frequencies are small compared to the
ion-neutral collision frequencies,%
\begin{equation}
\left\vert \omega\right\vert ,\ kV_{\max},|\omega_{Ds}|\ll\nu_{jn}\ll\nu
_{en},\qquad k^{2}\lambda_{Ds}^{2}\ll1. \label{long_conditions}%
\end{equation}
Here $V_{\max}$ is the largest between the mean flow speeds, $U_{j}=|\vec{U}_{j}|$, and ion
thermal speeds, $(V_{\mathrm{Th}})_{j}=(T_{j}/m_{j})^{1/2}$.

We give special attention to the LWL for three major reasons:

\begin{enumerate}
\item The minimum threshold for all collisional plasma instabilities is
usually reached within the LWL. If at a given location in space there is no
linear instability within the LWL then this location is linearly stable for
all $\omega$,$\vec{k}$-waves.

\item As we mentioned above, fluid-model Eqs.~(\ref{my_fluid_equations_cont}%
)--(\ref{my_fluid_equations_temperat}) are strictly valid only within the LWL.
Outside this limit, a stabilizing effect of ion Landau damping becomes
crucial, so that the rigorous treatment requires employing there a more
physically consistent kinetic approach.

\item In the LWL, all different instability-driving mechanisms are linearly
separated (see below). This makes the analysis of each physical mechanism much easier.
\end{enumerate}

One can easily verify that in the LWL the absolute values of $A_{s}$,
$B_{e,j}$ (but not the ratio $A_{j}/A_{e}$) are automatically small. To the
first-order accuracy with respect to the small quantities
\begin{equation}
\left\vert A_{s}\right\vert ,\left\vert B_{s}\right\vert ,\frac{|\omega_{Ds}%
|}{\nu_{sn}},\frac{k_{\parallel}^{2}}{k_{\perp}^{2}},k^{2}\lambda_{Ds}^{2}%
\ll1,\label{small}%
\end{equation}
from Eq.~(\ref{phi=snova}) we have
\[
N_{s}\approx\left(  1+A_{s}+\frac{2A_{s}+3B_{s}}{3\mu_{s}}\right)  A_{s},
\]
so that general dispersion Eq.~(\ref{disperga_snova}) reduces to%
\begin{widetext}
\begin{align}
&D(\omega,\vec{k})
\equiv1+\sum_{j=1}^{p}\frac{\rho_{j}\alpha_{j}A_{j}}{A_{e}}\left(
1+A_{j}-A_{e}+\frac{2A_{j}+3B_{j}}{3\mu_{j}}-\frac{2A_{e}+3B_{e}}{3\mu_{e}}\right)
\nonumber \\
&-~\frac{k^{2}\lambda_{De}^{2}}{A_{e}}\left(  1-A_{e}-\frac{2A_{e}+3B_{e}%
}{3\mu_{e}}\right)  = 0,\label{D_reduced}%
\end{align}
\end{widetext}
where $A_{j,e}$ and $B_{j,e}$ are given by Eqs.~(\ref{A_e,B_e}%
)--(\ref{ion_angles}) and $\mu_{s}$ are defined in Eq.~(\ref{ABmu}).

Reduced Eq.~(\ref{D_reduced}) has certain advantages over general
Eq.~(\ref{disperga_snova}). First, in the LWL the quantity $|\operatorname{Im}%
D(\omega,\vec{k})|$ turns out to be automatically small compared to
$|\operatorname{Re}D(\omega,\vec{k})|$, as well as the growth/damping rate,
$|\gamma|$, becomes automatically small compared to the real wave frequency,
$\omega_{r}$. This allows one to treat the wave phase-velocity relation for
$\omega_{r}(\vec{k})$ (the \textquotedblleft zeroth-order\textquotedblright%
\ approximation)\ separately from the instability driving (the
\textquotedblleft first-order\textquotedblright\ approximation).\ Second, as
we already mentioned, Eq.~(\ref{D_reduced}) allows one to explicitly separate
all instability driving mechanisms and diffusion losses, making the
instability analysis much easier.

Under condition of $|\gamma|\ll\omega_{r}$, if we also neglect all first-order
small terms in the RHS\ of Eq.~(\ref{D_reduced}) and use $\omega\approx
\omega_{r}$ in the highest-order terms, $D(\omega,\vec{k})\approx
\operatorname{Re}D(\omega_{r},\vec{k})=D_{0}(\omega_{r},\vec{k})$, we
obtain the equation for $\omega_{r}(\vec{k})$. Real solutions of $D_{0}%
(\omega_{r},\vec{k})=0$ will provide the zeroth-order phase-velocity relations
for the linear harmonic waves. To the next-order approximation, adding the
small imaginary parts and solving the first-order equation with $i\gamma$,
included in the complex wave frequency $\omega$, we obtain an approximate
expression for the growth/damping rate,%
\begin{equation}
\gamma\approx-\left.  \frac{\operatorname{Im}D(\omega,\vec{k})}{\partial
D_{0}(\omega,\vec{k})/\partial\omega}\right\vert _{\omega=\omega_{r}}.
\label{gamma_small}%
\end{equation}

Below we implement all these procedures. In
Sec.~\ref{Wave phase velocity relation}, we discuss the zeroth-order
approximation for the dominant real part of the Doppler-shifted wave frequency
$\omega_{De}=\omega-\vec{k}\cdot\vec{V}_{0}$. This real part is responsible
for the wave phase-velocity relation. For arbitrarily magnetized multi-species
ions, the explicit analytical solutions for $\omega_{De}\approx
\operatorname{Re}\omega_{De}$ can be found only in some particular cases.
Bearing in mind the actual physical conditions (especially in solar
chromosphere), we find approximate solutions that have fairly broad field of
applicability. In Sec.~\ref{Instability growth rates: physical mechanisms}, we
find the explicit expressions for the instability growth rates for each
component of the TFBI and damping mechanisms in terms of $\omega_{De}$.
Section~\ref{Threshold electric field}, discusses the major result of the
linear theory. i.e., the instability threshold. We obtain the general
expression for the threshold electric field $\vec{E}_{\mathrm{Thr}}$ (or the
corresponding $\vec{E}_{\mathrm{Thr}}\times\vec{B}_{0}$ speed) and discuss
particular cases.

\subsection{Zeroth-order approximation: wave phase-velocity
relation\label{Wave phase velocity relation}}

The zeroth-order relation for the dominant real part of the wave frequency is
obtained by neglecting in the RHS\ of Eq.~(\ref{D_reduced}) all terms
proportional to $A_{s}$ and $B_{s}$, except their ratio $A_{j}/A_{e}$. This
yields the following equation:
\begin{align}
&  D(\omega,\vec{k})\approx D_{0}(\omega_{r},\vec{k})=\left.  1+\sum_{j=1}%
^{p}\operatorname{Re}\left(  \frac{\rho_{j}\alpha_{j}A_{j}}{A_{e}}\right)
\right\vert _{\omega=\omega_{r}}\nonumber\\
&  =1+\omega_{De}\sum_{j=1}^{p}\frac{\rho_{j}}{(1+\kappa_{j}^{2})(\Omega
_{e}+\vec{k}\cdot\vec{U}_{j})\psi_{j}}=0, \label{D_0}%
\end{align}
where $\rho_{j}=(q_{j}/e)(n_{j0}/n_{e0})$,
\begin{equation}
\psi_{j}\equiv\frac{1}{\kappa_{e}\kappa_{j}}\left(  1+\frac{\kappa_{e}%
^{2}k_{\parallel}^{2}}{k_{\perp}^{2}}\right)  , \label{psi_j}%
\end{equation}
and by $\omega_{De}=\omega-\vec{k}\cdot\vec{V}_{0}$ we imply here and
throughout the remainder of the text the dominant real part of the Doppler-shifted wave
frequency, $\omega_{De}\approx\omega_{r}-\vec{k}\cdot\vec{V}_{0}$.

For the particular case of single-species ions (SSIs, $j\rightarrow i$),
Eq.~(\ref{D_0}) reduces to a much simpler equation, $1+\omega_{De}%
/[(1+\kappa_{i}^{2})(\omega_{De}+\vec{k}\cdot\vec{U}_{i})\psi_{i}]=0$,
yielding%
\begin{align}
\omega_{De}  &  =\omega_{r}-\vec{k}\cdot\vec{V}_{0}=-\ \frac{\left(
1+\kappa_{i}^{2}\right)  (\vec{k}\cdot\vec{U}_{i})\psi_{i}}{1+\left(
1+\kappa_{i}^{2}\right)  \psi_{i}},\nonumber\\
\omega_{Di}  &  =\omega_{De}+\vec{k}\cdot\vec{U}_{i}=\frac{\vec{k}\cdot\vec
{U}_{i}}{1+\left(  1+\kappa_{i}^{2}\right)  \psi_{i}},\nonumber\\
\omega_{r}  &  =\frac{\vec{k}\cdot\lbrack\vec{V}_{0}+\left(  1+\kappa_{i}%
^{2}\right)  \psi_{i}\vec{V}_{i0}]}{1+\left(  1+\kappa_{i}^{2}\right)
\psi_{i}}, \label{ss_phase_velocity}%
\end{align}
in full agreement with the previously published results, see, e.g.,
Refs.\ \onlinecite{Dimant:Model03,Dimant:Ion04} and references therein. For the linearly
unstable waves with $\vec{k}\cdot\vec{U}_{i}>0$, the Doppler-shifted wave
frequency in the electron-fluid frame of reference, $\omega_{De}$, is
negative, while the Doppler-shifted wave frequency in the ion-fluid frame,
$\omega_{Di}$, is positive. Physically, this means that electrons move somewhat ahead of
the wave, while ions lag behind it. This feature is important for the
self-consistent formation of the long-lived compression/rarefaction waves,
which in low-ionized highly dissipative plasmas can only be sustained by an
external DC\ electric field $\vec{E}_{0}$.

The solution of Eq.~(\ref{D_0}) simplifies dramatically also in the case of
unmagnetized multi-species ions, $\kappa_{j}\ll1$.
If all ions are essentially unmagnetized (as, e.g., in the E-region ionosphere
at altitudes below 115~km and, perhaps, at some cold regions in the
mid-chromosphere of the quiet sun) then all relative $e$-$i$ velocities are
almost equal, $\vec{U}_{j}\approx\vec{V}_{0}=\vec{E}_{0}\times\hat{b}/B_{0}$.
In this case, all ion Doppler-shited frequencies $\omega_{Dj}$ are shifted
from $\omega_{De}$ approximately by the same $\vec{k}$-dependent quantity
$\vec{k}\cdot\vec{V}_{0}$,%
\begin{equation}
\omega_{Dj}\approx\omega_{Di}\equiv\omega_{De}+\vec{k}\cdot\vec{V}_{0}.
\label{Omega_i_equal}%
\end{equation}
This reduces general Eq.~(\ref{D_0}) to an easily solvable equation
\begin{equation}
1+\frac{\omega_{De}}{\omega_{De}+\vec{k}\cdot\vec{V}_{0}}\sum_{j=1}^{p}%
\frac{\rho_{j}}{\psi_{j}}=0. \label{Phase_velocity_simple}%
\end{equation}
This means that all different $p$ roots of Eq.~(\ref{D_0}) degenerate into a
single root for $\omega_{De}$, with all $\omega_{Dj}$ equal to the same common
value for all ions, $\omega_{Di}$,
\begin{equation}
\omega_{De}=-\ \frac{(\vec{k}\cdot\vec{V}_{0})\Psi}{1+\Psi},\qquad\omega
_{Di}=\frac{\vec{k}\cdot\vec{V}_{0}}{1+\Psi}, \label{Omegas_simple}%
\end{equation}
where the parameter%
\begin{equation}
\Psi\equiv\left(  \sum_{j=1}^{p}\frac{\rho_{j}}{\psi_{j}}\right)  ^{-1}
\label{Psi}%
\end{equation}
generalizes the parameter $\psi_{j}=\psi_{i}$ in the standard SSI solution
(since $\sum_{j=1}^{p}\rho_{j}=1$, in the SSI case $\Psi=\psi_{i}$).

Before looking at more general cases, it is useful to rewrite, in accord with
Eqs.~(\ref{UU}) and (\ref{ion_angles}), the scalar product $\vec{k}\cdot
\vec{U}_{j}$ as%
\begin{equation}
\vec{k}\cdot\vec{U}_{j}=G_{j}kV_{0},\qquad G_{j}\equiv\frac{\cos\chi_{j}%
}{(1+\kappa_{j}^{2})^{1/2}}=\frac{\cos\theta-\kappa_{j}\sin\theta}%
{1+\kappa_{j}^{2}}. \label{kU_via_G}%
\end{equation}
where the dimensionless parameter $G_{j}$ is independent of $k$ and $V_{0}$.
Accordingly, the electron Doppler-shifted frequency, $\omega_{De}$, as a
solution of Eq.~(\ref{D_0}), and hence $\omega_{Dj}=\omega_{De}+\vec{k}%
\cdot\vec{U}_{j}$, should be similarly written in proportion to $kV_{0}$,
\begin{equation}
\omega_{De}=\zeta_{e}kV_{0},\qquad\omega_{Dj}=\zeta_{j}kV_{0},\qquad\zeta
_{j}=\zeta_{e}+G_{j}. \label{Omega_e_via_zeta_e}%
\end{equation}
As a result, Eq.~(\ref{D_0}) reduces to an equation for the dimensionless
variable $\zeta_{e}$,%
\begin{equation}
1+\zeta_{e}\sum_{j=1}^{p}\frac{\rho_{j}}{(1+\kappa_{j}^{2})(\zeta_{e}%
+G_{j})\psi_{j}}=0, \label{Equation_for_zeta_e}%
\end{equation}
that involves neither $k$ nor $V_{0}$. This equation depends only on the
$\vec{k}$-direction (via $\theta$) and local magnetization parameters
$\kappa_{j}$, $\psi_{j}$.

In the general case of multi-species ions with different $\vec{k}\cdot\vec
{U}_{j}$ (i.e., with different $G_{j}$), Eq.~(\ref{D_0}) can be reduced to a
polynomial equation of degree $p$, where $p$ is the total number of the ion
species. For arbitrary $p$, this equation is either analytically unsolvable
(for $p\geq5$) or has cumbersome exact solutions (for $p=2,3,4$). Apart from
degenerate cases, Eq.~(\ref{D_0}) has exactly $p$ real negative roots for
$\omega_{De}=\zeta_{e}kV_{0}$, while all corresponding $\omega_{Dj}=\zeta
_{j}kV_{0}$ are positive.

\begin{figure}
\noindent\includegraphics[width=0.5\textwidth]{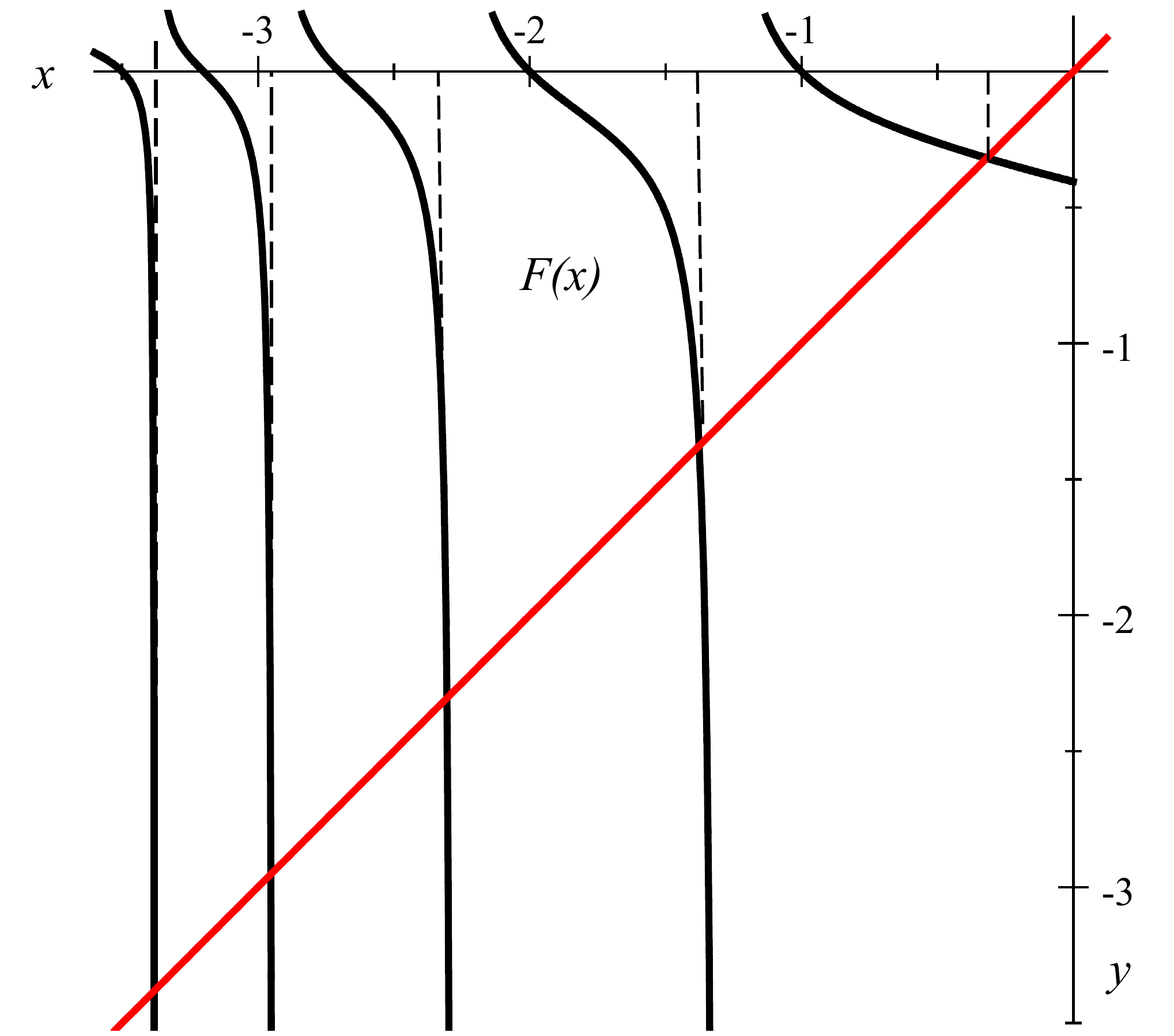}
\caption{\label{Solution_Fx}An example of the graphic solution of Eq.~(\ref{x=F(x)}). Solid curves show $p$ isolated segments of
$y=F(x)$, where $p$ vertical dashed lines mark $x=b_{j}$. All $p$ solutions of Eq.~(\ref{x=F(x)})
correspond to the intersections of the solid curves with the diagonal red line $y=x$. The total number of ion species ($p=5$)
and the specific values of $a_{j}$ used in this example serve only to illustrate the general behavior of the solutions;
they do not correspond to any real physical situation in the solar
chromosphere or elsewhere. }
\end{figure}

To illustrate the latter statement, it is useful to rewrite
Eq.~(\ref{Equation_for_zeta_e})\ as
\begin{equation}
\zeta_{e}=F(\zeta_{e}),\label{x=F(x)}%
\end{equation}
where%
\begin{widetext}
\begin{equation}
F(\zeta_{e})\equiv-\left(  \sum_{j=1}^{p}\frac{\xi_{j}}{\zeta_{e}-a_{j}%
}\right)  ^{-1},\qquad\xi_{j}\equiv\frac{\rho_{j}}{(1+\kappa_{j}^{2})\psi_{j}%
},\qquad a_{j}=-G_{j}.\label{F,mu_a}%
\end{equation}
\end{widetext}
Figure \ref{Solution_Fx} shows schematically the two sides of Eq.~(\ref{x=F(x)}) for a
generic set of different $\xi_{j}$ and $a_{j}$.
All $p$ roots of $\zeta_{e}=F(\zeta_{e})$ are given by the intersections of
the diagonal $y=\zeta_{e}$ with the curve $y=F(\zeta_{e})$. For any integer
$p>1$, the entire curve $y=F(\zeta_{e})$ represents $p$ isolated segments
$y=F_{s}(\zeta_{e})$, separated by $p-1$ singularities of the $1/(\zeta
_{e}-b_{s})$-kind (bear in mind that $b_{s}\neq a_{s}$). The vertical values
of each segment boundary span the entire $(-\infty,\infty)$ range of the
$y$-value, either in semi-infinite $\zeta_{e}$ domains (for the two edge
segments) or within finite domains between two adjacent singularities. Each
singularity, $\zeta_{e}=b_{s}$, in turn, is situated between two adjacent
zeroes of $F(\zeta_{e})$, $(\zeta_{e})_{s}=a_{s}$ and $(\zeta_{e})_{s+1}=a_{s+1}$.
All $p$ zeroes of $F(\zeta_{e})$, $(\zeta_{e})_{s}=a_{s}$, as well as all
$p-1$ singularities, $(\zeta_{e})_{s}=b_{s}$, are negative. This pertains to
all $p$ roots $\zeta_{e}$ of equivalent Eqs.~(\ref{Equation_for_zeta_e}) and
(\ref{x=F(x)}).

Thus, if all $\vec{k}\cdot\vec{U}_{j}=G_{j}kV_{0}$ are different then the
solution of Eq.~(\ref{x=F(x)}) has exactly $p$ negative roots of $\omega_{De}$.
In the general case, these roots can be found numerically. Each root
corresponds to a separate wave mode. However, we will be interested only in
one solution that corresponds to the minimum instability threshold field (if
there are more than one linearly unstable modes). Based on particular cases
described below, we may suppose that this solution has the minimum value of
$|\zeta_{e}|$ corresponding to the largest values of $\zeta_{j}=\zeta
_{e}+G_{j}$.

Now we consider particular cases that will allow us to obtain explicit
analytic solutions. First, if all ions are essentially unmagnetized
($\kappa_{j}\ll1$, see above) then all $G_{j}\approx\cos\theta$, so that
Eq.~(\ref{Equation_for_zeta_e}) reduces to $1+\zeta_{e}/[(\zeta_{e}+\cos
\theta)\Psi]=0$ with the obvious solution
\[
\zeta_{e}=-\ \frac{\Psi\cos\theta}{1+\Psi},\qquad\zeta_{j}=\frac{\cos\theta
}{1+\Psi},
\]
where $\Psi$ is defined by Eq.~(\ref{Psi}). This solution is equivalent to
Eq.~(\ref{Omegas_simple}). However, if at least one ion species is partially magnetized,
$\kappa_{j}\gtrsim1$, then the situation is less simple.

As a second particular case, we consider partially magnetized ion
species, assuming first that $\kappa_{j}\gtrsim1$ holds for all ions (more accurate
conditions will be discussed below). For partially magnetized ions, the
quantities $\vec{k}\cdot\vec{V}_{j0}$ are not negligibly small. Being unable
to find the general exact solution of Eq.~(\ref{Equation_for_zeta_e}) or
(\ref{x=F(x)}), one can utilize an approximate approach, implemented
earlier for the pure FBI \cite{Fletcher:Effects18}. This approach is based on
the existence of a small parameter%
\begin{equation}
\Theta_{j}\equiv\sqrt{\frac{\kappa_{j}}{\kappa_{e}}}=\sqrt{\frac{m_{e}\nu
_{en}}{m_{j}\nu_{jn}}}. \label{Theta_j}%
\end{equation}
For example, throughout the E-region ionosphere, $\Theta_{j}=\Theta_{0}%
\simeq1.4\times10^{-2}$, see Refs.\ \onlinecite{Dimant:Model03,Dimant:Ion04}. In the solar
chromosphere, dominated by ion collisions with the light atomic hydrogen, the
values of $\Theta_{j}$ are typically larger (see below), but they always obey
a slightly weaker inequality, $\Theta_{j}^{2}\ll1$.

Fletcher et al. \cite{Fletcher:Effects18} used the following idea. Restricting
the treatment to strictly perpendicular waves, $k_{\parallel}=0$, for which we
usually expect the minimal threshold field, one can write the parameter
$\psi_{j}$ defined by Eq.~(\ref{psi_j}) as $\psi_{j}=\Theta_{j}^{2}/\kappa
_{j}^{2}$. Then for partially magnetized ion species, assuming $\kappa_{j}^{2}
\gg\Theta_{j}^{2}$, one automatically has $\psi_{j}\ll1$. In the
E-region ionosphere, at altitudes where $\psi_{j}=\psi\ll1$ (usually, above
100 km of altitude), this automatically provides $|\zeta_{e}|\ll1$. Expecting a
similar inequality to hold for all multi-species ions in other media, one can
easily solve Eq.~(\ref{Equation_for_zeta_e}) by neglecting $|\zeta_{e}|$
compared to $G_{j}$ in all denominators. This reduces the
original high-order polynomial equation to a linear one with the simple (and
unique) solution,
\begin{subequations}
\label{unique_solution}%
\begin{align}
\zeta_{e}  &  \approx-\left[  \sum_{j=1}^{p}\frac{\rho_{j}}{(\cos\theta
-\kappa_{j}\sin\theta)\psi_{j}}\right]  ^{-1} \nonumber\\
&  =-\left[  \sum_{j=1}^{p}\frac{\rho_{j}}{(1+\kappa_{j}^{2})^{1/2}\psi
_{j}\cos\chi_{j}}\right]  ^{-1}, \label{unique_solution_Omega_e}\\
\zeta_{j}  &  \approx G_{j}=\frac{\cos\theta-\kappa_{j}\sin\theta}%
{1+\kappa_{j}^{2}}, \label{unique_solution_Omega_j}%
\end{align}
for each ion species $j$. The condition for this approximate solution,
$|\zeta_{e}|\ll|G_{j}|$, requires%
\end{subequations}
\begin{equation}
\left\vert \left[  \sum_{j=1}^{p}\frac{\rho_{j}}{(\cos\theta-\kappa_{j}%
\sin\theta)\psi_{j}}\right]  ^{-1}\right\vert \ll\frac{|\cos\theta-\kappa
_{j}\sin\theta|}{1+\kappa_{j}^{2}}. \label{requires}%
\end{equation}
Assuming both $|\cos\theta-\kappa_{j}\sin\theta|$ and $\kappa_{j}^{2}$ to be
of order unity, we reduce Eq.~(\ref{requires}) to a much simpler criterion:
$\Psi\ll1$. If the wave direction is such that for some specific ion species
the flow angle $\theta$ is close to $\tan^{-1}\kappa_{j}$ (leading to
$|\cos\theta-\kappa_{j}\sin\theta|\rightarrow0$) then the corresponding
contribution to the summation, $j=i$, dominates, reducing the
Eq.~(\ref{requires}) to%
\begin{equation}
\psi_{j}=\frac{\Theta_{j}^{2}}{\kappa_{j}^{2}}\ll1. \label{psi_j<<1}%
\end{equation}

The above two cases of low-magnetized ions, $\kappa_{j}\ll1$ (equivalent to
$\psi_{j}\gg\Theta_{j}^{2}$) and the low-$\psi_{j}$ case, $\psi_{j}\ll1$
(equivalent to $\kappa_{j}^{2}\gg\Theta_{j}^{2}$) overlap under fairly broad
conditions of $\Theta_{j}^{2}\ll\psi_{j}\ll1$, equivalent to $1\gg\kappa
_{j}^{2}\gg\Theta_{j}^{2}$. These two overlapping cases together cover a
significant domain of the collisional plasma parameters, but they still do not
encompass all possible situations. The reason is that the relevant
ion-magnetization conditions were imposed for all ions. However, there is a
possibility that at a given location the conditions $\kappa_{j}\ll1$ and
$\kappa_{j}\gtrsim1$ are satisfied separately for different ion species. In
those cases, Eq.~(\ref{Equation_for_zeta_e}) does not necessarily reduce to a
simple linear equation for $\zeta_{e}$. In some cases, if the ratios $\rho
_{j}/\psi_{j}$ with small $\psi_{j}\ll1$ dominate over all the others with
$\psi_{j}\gtrsim1$ then this case can be approximately reduced to the above
low-$\psi_{j}$ case. If, however, the corresponding ion concentrations
$\rho_{j}$ are too small, $\rho_{j}\lesssim\psi_{j}$, then the situation is
more complicated.

For the solar chromosphere, however, the general situation simplifies
dramatically if we assume that for both $e$-$n$ and $i$-$n$ collisions the MMC
approximation holds (see Sec.~\ref{Mean particle flows}). In this approximation, for
elastic $i$-$n$ or $e$-$n$ collisions (assuming first no charge exchange
between the colliding ions and atoms of different materials), the expression
for the $s$-$n$ collision frequency is given by \cite{Schunk:Ionospheres09,Oppenheim:Newly20},
\begin{widetext}
\begin{equation}
\nu_{sn}=\frac{2.21\pi n_{n}m_{n}}{m_{s}+m_{n}}\sqrt{\frac{\alpha_{n}e^{2}%
}{4\pi\epsilon_{0}\mu_{sn}}}\approx1.96n_{n}\sqrt{\frac{\alpha_{n}e^{2}m_{n}%
}{\epsilon_{0}m_{s}\left(  m_{s}+m_{n}\right)  }}, \label{MMC}%
\end{equation}
\end{widetext}
where $\mu_{sn}=m_{s}m_{n}/(m_{s}+m_{n})$ is the reduced mass of the two
colliding particles, $n_{n}$ is the neutral particle density, $\epsilon_{0}$
is the permittivity of free space, and $\alpha_{n}$ is the neutral-particle
polarizability. In the solar chromosphere, the dominant neutral component\ is
the atomic hydrogen (H) for which we have $\alpha_{n}\approx\alpha
_{\mathrm{H}}\approx0.67\times10^{-24}%
\operatorname{cm}%
^{3}$, see Ref.\ \onlinecite{Schunk:Ionospheres09}.

Elastic-collision Eq.~(\ref{MMC}) applies there only to $i$-H collisions of
heavy ions like C$^{+}$, Mg$^{+}$, Fe$^{+}$, etc. ($s=j^{+}\neq\mathrm{H}^{+}%
$), whose mass is significantly larger than the atomic mass of the neutral
collision partner H ($m_{n}=m_{\mathrm{H}}$; recall that here we ignore any
contribution of He). For these heavy ions, one can neglect the hydrogen mass
$m_{\mathrm{H}}$ compared to $m_{j^{+}}$, so that $\mu_{j^{+}\mathrm{H}%
}\approx m_{\mathrm{H}}$ and
\begin{equation}
\nu_{j^{+}\mathrm{H}}\approx1.96n_{\mathrm{H}}\sqrt{\frac{\alpha_{\mathrm{H}%
}e^{2}m_{\mathrm{H}}}{\epsilon_{0}m_{j^{+}}^{2}}}\approx2.11\times
10^{5}\ \frac{m_{\mathrm{H}}}{m_{j^{+}}}\left(  \frac{n_{\mathrm{H}}}{10^{20}%
\operatorname{m}%
^{-3}}\right)
\operatorname{s}%
^{-1}. \label{nu_jH}%
\end{equation}
The inverse proportionality of $\nu_{j^{+}\mathrm{H}}$ to the ion mass
directly follows from the fact that heavy chromospheric ions collide
predominantly with the much lighter neutral atoms (H).

For the H$^{+}$-H collisions, to a reasonable accuracy, one can also use the
MMC approximation, i.e., assume nearly constant $\nu_{\mathrm{H}^{+}%
\mathrm{H}}$, but not the specific elastic-collision expression given by
Eq.~(\ref{MMC}). Using Figure 1 from Ref.\ \onlinecite{Vranjes:Collisions13} (after smoothing the
corresponding curve over frequent oscillations), we approximately obtain%
\begin{equation}
\nu_{\mathrm{H}^{+}\mathrm{H}}\simeq2\times10^{6}\left(  \frac{n_{\mathrm{H}}%
}{10^{20}%
\operatorname{m}%
^{-3}}\right)
\operatorname{s}%
^{-1}. \label{nu_H^+H}%
\end{equation}
Note that Eq.~(\ref{MMC}) would result in about twenty times smaller value for
$\nu_{\mathrm{H}^{+}\mathrm{H}}$. The charge-exchange process is the major
reason for the much higher total $\mathrm{H}^{+}$-$\mathrm{H}$ collision frequency.

For the e-H collisions, using Eq.~(\ref{MMC}), we obtain:%
\begin{equation}
\nu_{e\mathrm{H}}\approx1.96n_{n}\sqrt{\frac{\alpha_{n}e^{2}}{\epsilon
_{0}m_{e}}}\approx0.905\times10^{7}\left(  \frac{n_{\mathrm{H}}}{10^{20}%
\operatorname{m}%
^{-3}}\right)
\operatorname{s}%
^{-1}. \label{nu_eH}%
\end{equation}
Figure~4 from Ref.\ \onlinecite{Vranjes:Collisions13} provides a value of $\nu_{e\mathrm{H}}$
reasonably close to this.

The fact that the collision frequency $\nu_{j^{+}\mathrm{H}}$ for $j^{+}%
\neq\mathrm{H}^{+}$ is inversely proportional to the ion mass means that the
magnetization ratio $\kappa_{j^{+}}=\Omega_{j^{+}}/\nu_{j^{+}\mathrm{H}}$ has approximately
the same common value for all heavy-ion collisions with the neutral hydrogen,
\begin{equation}
\kappa_{i}=\kappa_{j^{+}}\approx\frac{0.51B_{0}}{n_{\mathrm{H}}}\sqrt
{\frac{\epsilon_{0}}{\alpha_{\mathrm{H}}m_{\mathrm{H}}}}\approx0.45\left(
\frac{B_{0}}{10%
\operatorname{G}%
}\right)  \left(  \frac{10^{20}%
\operatorname{m}%
^{-3}}{n_{\mathrm{H}}}\right)  . \label{kappa_j}%
\end{equation}
Due to this, for all heavy ions with $m_{j^{+}}\gg m_{\mathrm{H}}$, we have
equal values of the parameter%
\[
\psi_{j^{+}0}=\frac{1}{\kappa_{e}\kappa_{j^{+}}}=\frac{\Theta_{i}^{2}}%
{\kappa_{i}^{2}},
\]
where%
\begin{equation}
\Theta_{i}\equiv\Theta_{j^{+}\neq\mathrm{H}^{+}}=\sqrt{\frac{\kappa_{j^{+}}%
}{\kappa_{e}}}=\sqrt{\frac{m_{e}\nu_{e\mathrm{H}}}{m_{j^{+}}\nu_{j^{+}%
\mathrm{H}}}} \label{Theta_j+}%
\end{equation}
with the subscript $i$ applying only to the heavy ions. For these ions, the
parameter $\Theta_{i}^{2}$ is fairly small,%
\begin{equation}
\Theta_{i}^{2}\approx\sqrt{\frac{m_{e}}{m_{\mathrm{H}}}}\approx2.334\times
10^{-2}. \label{Theta^2_heavy}%
\end{equation}

For the $\mathrm{H}^{+}$-$\mathrm{H}$ collision magnetization parameter, we
obtain%
\begin{equation}
\kappa_{\mathrm{H}^{+}}\approx4.79\times10^{-2}\left(  \frac{B_{0}}{10%
\operatorname{G}%
}\right)  \left(  \frac{10^{20}%
\operatorname{m}%
^{-3}}{n_{\mathrm{H}}}\right)  . \label{kappa_H}%
\end{equation}
This value is an order of magnitude smaller than $\kappa_{i}=\kappa_{j^{+}}$.
Accordingly, $\Theta_{\mathrm{H}^{+}}^{2}$ turns out to be an order of
magnitude smaller than $\Theta_{i}^{2}$,
\begin{equation}
\Theta_{\mathrm{H}^{+}}^{2}=\sqrt{\frac{\kappa_{\mathrm{H}^{+}}}{\kappa_{e}}%
}\approx2.4646\times10^{-3}. \label{Theta^2_H+}%
\end{equation}
We will use the smallness of the parameters $\Theta_{i}^{2}$ and
$\Theta_{\mathrm{H}^{+}}^{2}$ below.

Thus, instead of $p$ totally different values of ion magnetization parameters,
under conditions of $m_{j}\gg m_{\mathrm{H}}$ we have only two distinct values
of the ion magnetization parameter: $\kappa_{i}$ for all heavy ions and
$\kappa_{\mathrm{H}^{+}}$ for H$^{+}$. As a result,
Eq.~(\ref{Equation_for_zeta_e}) reduces to a much simpler equation:
\begin{equation}
1+\frac{\zeta_{e}\xi_{\mathrm{H}^{+}}}{\zeta_{e}+G_{\mathrm{H}^{+}}}%
+\frac{\zeta_{e}\xi_{i}}{\zeta_{e}+G_{j}}=0, \label{quadratic}%
\end{equation}
where, in accord with Eqs.~(\ref{VVU_abs_values}) and (\ref{ion_angles}),
\begin{align}
&  \xi_{\mathrm{H}^{+}}\equiv\frac{\rho_{\mathrm{H}^{+}}\kappa_{\mathrm{H}%
^{+}}^{2}}{(1+\kappa_{\mathrm{H}^{+}}^{2})\Theta_{\mathrm{H}^{+}}^{2}}%
=\frac{\varepsilon\rho_{\mathrm{H}^{+}}\kappa_{i}^{2}}{(1+\varepsilon
^{2}\kappa_{i}^{2})\Theta_{i}^{2}},\nonumber\\
&  \xi_{i}\equiv\frac{\rho_{i}\kappa_{i}^{2}}{(1+\kappa_{i}^{2})\Theta_{i}%
^{2}},\qquad\rho_{i}\equiv\sum_{i^{+}\neq\mathrm{H}^{+}}\rho_{i^{+}}%
=1-\rho_{\mathrm{H}^{+}}.\nonumber\\
&  G_{i}\equiv\frac{\vec{k}\cdot\vec{U}_{i}}{kV_{0}}=\frac{\cos\theta
-\kappa_{i}\sin\theta}{1+\kappa_{i}^{2}},\nonumber\\
&  G_{\mathrm{H}^{+}}\equiv\frac{\vec{k}\cdot\vec{U}_{\mathrm{H}^{+}}}{kV_{0}%
}=\frac{\cos\theta-\varepsilon\kappa_{i}\sin\theta}{1+\varepsilon^{2}%
\kappa_{i}^{2}}. \label{Various}%
\end{align}
Here $\varepsilon$ is a small dimensionless parameter,
\begin{equation}
\varepsilon\equiv\frac{\kappa_{\mathrm{H}^{+}}}{\kappa_{i}}=\frac
{\Theta_{\mathrm{H}^{+}}^{2}}{\Theta_{i}^{2}}\approx0.1056. \label{epsilon}%
\end{equation}
According to Eq.~(\ref{Various}), given constant $k$, $\theta$, $V_{0}$,
$\rho_{\mathrm{H}^{+}}$ and the small parameters $\Theta_{i}^{2}$ and
$\varepsilon$ defined by Eqs.~(\ref{Theta^2_heavy}) and (\ref{epsilon}), all
remaining quantities in Eq.~(\ref{quadratic}) are expressed in terms of only
one parameter, $\kappa_{i}^{2}$, which varies with the total hydrogen density
and magnetic field according to Eq.~(\ref{kappa_j}).

In an obvious way, Eq.~(\ref{quadratic}) reduces to a quadratic equation for
$\zeta_{e}=\omega_{De}/(kV_{0})$,
\begin{equation}
(1+\xi_{\mathrm{H}^{+}}+\xi_{i})\zeta_{e}^{2}+[(1+\xi_{\mathrm{H}^{+}}%
)G_{i}+(1+\xi_{i})G_{\mathrm{H}^{+}}]\zeta_{e}+G_{\mathrm{H}^{+}}G_{i}=0,
\label{quadratic_for_z}%
\end{equation}
whose two exact roots, $\zeta_{e}^{(1,2)}$, can be written as
\begin{subequations}
\label{Omega_e_two_roots}%
\begin{align}
\zeta_{e}^{(1)}  &  =-\ \frac{2G_{\mathrm{H}^{+}}G_{i}}{(1+\xi
_{\mathrm{H}^{+}})G_{i}+(1+\xi_{i})G_{\mathrm{H}^{+}}+Z}%
\,,\label{Omega_e^+}\\
\zeta_{e}^{(2)}  &  =-\ \frac{(1+\xi_{\mathrm{H}^{+}})G_{i}+(1+\xi
_{i})G_{\mathrm{H}^{+}}+Z}{2(1+\xi_{i}+\xi_{\mathrm{H}^{+}})}\,,
\label{Omega_e^-}%
\end{align}
where%
\end{subequations}
\begin{equation}
Z=\sqrt{[(1+\xi_{\mathrm{H}^{+}})G_{i}-(1+\xi_{i})G_{\mathrm{H}^{+}%
}]^{2}+4\xi_{\mathrm{H}^{+}}\xi_{i}G_{\mathrm{H}^{+}}G_{i}}\,. \label{Z}%
\end{equation}
We have written the two roots of a quadratic equation in an unconventional,
but equivalent, form which makes perfectly clear that each solution for
$\zeta_{e}$ is real and negative. Besides, in the large-$\xi_{i,\mathrm{H}}$
limit (see below), the conventional form of the solution for $\zeta_{e}^{(1)}$ would result in
subtraction of two major terms, while Eq.~(\ref{Omega_e^+}) allows one
to avoid that.

The above exact solution of simplified Eq.~(\ref{quadratic_for_z}) remains
complicated for analysis. Below, using the specific parameter relations
found above, we will construct a much simpler, but still reasonably accurate,
approximate solution.

First, assuming $\kappa_{i}^{2}\ll1$, so that automatically $\kappa
_{\mathrm{H}^{+}}^{2}=\varepsilon^{2}\kappa_{i}^{2}\ll1$, we reduce this case
to the fully unmagnetized case described above. In the specific case of
$\vec{U}_{j}\approx\vec{U}_{\mathrm{H}^{+}}\approx\vec{V}_{0}$,
Eq.~(\ref{quadratic}) yields%
\begin{equation}
\omega_{De}\approx-\ \frac{\vec{k}\cdot\vec{V}_{0}}{1+\xi_{i}+\xi
_{\mathrm{H}^{+}}}. \label{Omega_e_Solar_unmagnetized}%
\end{equation}
For $\vec{U}_{j}\approx\vec{U}_{\mathrm{H}^{+}}\approx\vec{V}_{0}$, this
solution also follows from Eq.~(\ref{Omega_e^+}).

Now we consider a broader span of the ion magnetization parameters that includes
$\kappa_{i}^{2}\gtrsim1$. In this, more general, case, one should no longer
expect $\vec{U}_{j}\approx\vec{U}_{\mathrm{H}^{+}}\approx\vec{V}_{0}$, though
$|\vec{U}_{j}|$ and $|\vec{U}_{\mathrm{H}^{+}}|$ usually have comparable
values. Indeed, only for strongly magnetized ions, $\kappa_{i}^{2}\gg1$, while
$\varepsilon^{2}\kappa_{i}^{2}\lesssim1$, we would have $|\vec{U}_{j}|\ll
|\vec{U}_{\mathrm{H}^{+}}|\sim V_{0}$, but this case is of no interest to us
because the large-$\kappa_{i}^{2}$ is linearly stable, as discussed above. In
all other cases, we typically have $|\vec{U}_{j}|\sim|\vec{U}_{\mathrm{H}^{+}%
}|\sim V_{0}$. Assuming in Eq.~(\ref{quadratic}) $|\zeta_{e}|$ to be small
compared to $G_{j}\sim G_{\mathrm{H}^{+}}$ (the condition will be discussed
below) and neglecting $\zeta_{e}$ in both denominators, we obtain%
\begin{equation}
\zeta_{e}\approx-\left(  \frac{\xi_{i}}{G_{j}}+\frac{\xi_{\mathrm{H}^{+}}%
}{G_{\mathrm{H}^{+}}}\right)  ^{-1}. \label{Omega_low_psi_Solar}%
\end{equation}
From Eq.~(\ref{Omega_low_psi_Solar}), assuming $G_{\mathrm{H}^{+}}\sim G_{j}$,
we obtain that the presumed condition of $|\zeta_{e}|\ll G_{\mathrm{H}^{+}%
}\sim G_{j}$ requires $\xi_{i,\mathrm{H}^{+}}\gg1$. It can be easily verified
that the approximate solution given by~(\ref{Omega_low_psi_Solar}) follows
from Eq.~(\ref{Omega_e^+}) if one neglects the \textquotedblleft
unity\textquotedblright\ compared to both $\xi_{\mathrm{H}^{+}}$ and $\xi_{i}%
$. According to Eq.~(\ref{Various}), unless the fraction of heavy ions is too
small ($\rho_{i}\lesssim\Theta_{i}^{2}\simeq0.02$), the condition of $\xi
_{i}\sim\rho_{i}\kappa_{i}^{2}/\Theta_{i}^{2}\gg1$ is automatically fulfilled
for $\kappa_{i}^{2}\sim1$. Similarly, unless $\rho_{\mathrm{H}^{+}}$ is too
small ($\rho_{\mathrm{H}^{+}}\lesssim\Theta_{i}^{2}/\varepsilon\simeq0.2$),
the condition $\xi_{\mathrm{H}^{+}}\sim\varepsilon\rho_{\mathrm{H}^{+}}%
\kappa_{i}^{2}/\Theta_{i}^{2}\gg1$ is also automatically fulfilled for the
same range of $\kappa_{i}^{2}\sim1$. In principle, if $\rho_{\mathrm{H}^{+}%
}\lesssim0.2$ then $\xi_{\mathrm{H}^{+}}\lesssim1$, so that $1$ cannot be
dropped compared to $\xi_{\mathrm{H}^{+}}$. However, this does not really
matter since the corresponding second term, $\xi_{\mathrm{H}^{+}%
}/G_{\mathrm{H}^{+}}$, in Eq.~(\ref{Omega_low_psi_Solar}) is small in itself
(compared to the first term, $\xi_{i}/G_{j}$). The inaccuracy of this small
term is largely inconsequential.

The two approximate solutions given by Eqs.~(\ref{Omega_e_Solar_unmagnetized})
and (\ref{Omega_e_two_roots}) match within the overlap range of $\Theta
_{i}^{2}/\rho_{i}\ll\kappa_{i}^{2}\ll1$, where both conditions of
$G_{\mathrm{H}^{+}}\approx G_{j}\approx\cos\theta$ and $\xi_{i}\gg1$ are
fulfilled simultaneously. For the most interesting cases, one can construct an
interpolation between the two solutions, using the simple ansatz:
\begin{equation}
\zeta_{e}\approx-\left(  \frac{\alpha_{1}+\xi_{i}}{G_{j}}+\frac{1-\alpha
_{1}+\xi_{\mathrm{H}^{+}}}{G_{\mathrm{H}^{+}}}\right)  ^{-1},
\label{interpolation}%
\end{equation}
where the specific value of the numeric parameter $\alpha_{1}$ can be chosen
between 0 and 1. This simple interpolation works well mostly within the range
of flow angles $\theta$ between $-45^{\circ}$ (the optimal angle for the pure
ETI) and $0^{\circ}$ (the optimal angle for the pure FBI).

\begin{figure*}
\noindent\includegraphics[width=\textwidth]{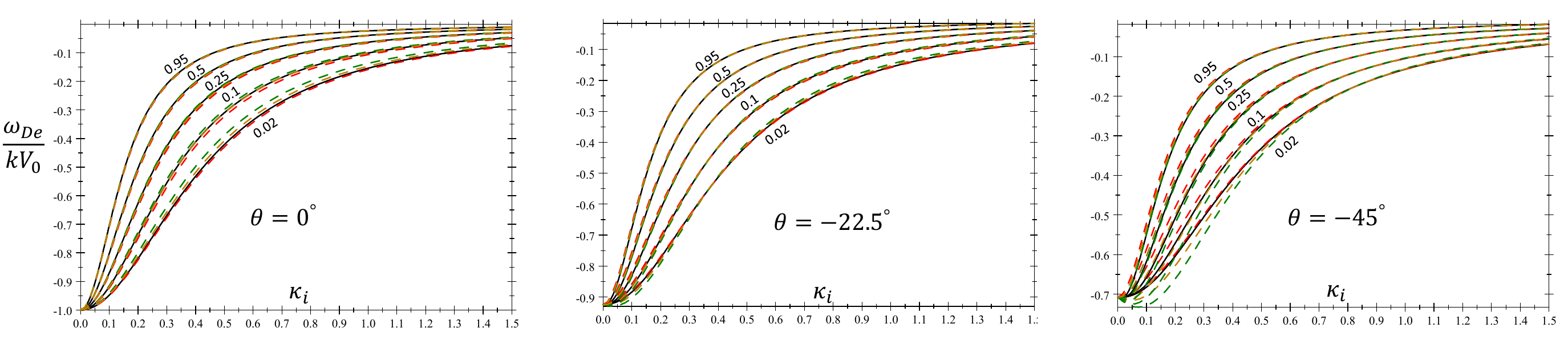}
\caption{\label{Theta_Figures}Solution of Eq.~(\ref{quadratic}) given by
Eq.~(\ref{Omega_e^+}) for three values of the flow angle $\theta$ (the solid curves) and for five different
values of the heavy-ion fraction, $\rho_{i}=1-\rho_{\mathrm{H}^{+}}$,
along with the corresponding interpolations given by Eq.~(\ref{interpolation}) and described
in the text (the dashed curves). In the interpolation curves, the red, yellow, and green
curves correspond to $\alpha_1$ equal to 0, 0.5, and 1, respectively.}
\end{figure*}
Figure\ \ref{Theta_Figures} shows the solution of Eq.~(\ref{quadratic}) given by
Eq.~(\ref{Omega_e^+}) for three values of the flow angle $\theta$. This
solution (normalized to $kV_{0}$) is shown by solid curves for five different
values of the heavy-ion fraction, $\rho_{i}=1-\rho_{\mathrm{H}^{+}}$ (shown
near the curves). Around these curves, there also interpolations given by
Eq.~(\ref{interpolation}) (shown by the dashed color curves) for three
different values of the fitting parameter $\alpha_{1}$ ($\alpha_{1}=0$, $0.5$,
$1$). For $\theta=-0^{\circ}$ and $\theta=-22.5^{\circ}$, the ansatz of
Eq.~(\ref{interpolation}) works reasonably well with any values of $\alpha
_{1}$, so that for $\rho_{i}\gtrsim0.25$, the interpolations are almost
indistinguishable from the exact solution. For $-22.5^{\circ}\lesssim
\theta\leq0^{\circ}$, the interpolation works reasonably well for all values
$\rho_{i}$, even for $\rho_{i}$ as low as $0.02$. For $\theta=-45^{\circ}$,
the interpolation starts deviating from the exact solution, though the
specific value of $\alpha_{1}$ matters only for low concentrations, $\rho
_{i}<0.1$, and mostly for low-magnetized ions, $\kappa_{i}<0.5$. Generally,
for most interesting cases of $\theta$ within $-45^{\circ}$ to $0^{\circ}$
range, the choice of $\alpha_{1}=1-\alpha_{1}=0.5$ seems to be optimal. For
all these cases, Eq.~(\ref{interpolation}) can serve as a reasonably accurate
and a more practical alternative to the cumbersome exact solution given by
Eq.~(\ref{Omega_e_two_roots}). Unfortunately, for angles beyond the domain of
$-45^{\circ}\lesssim\theta\leq0^{\circ}$, the simple interpolation of
Eq.~(\ref{interpolation}) often does not work well, so that one needs to apply
there the full solution given by Eq.~(\ref{Omega_e^+}).

In this analysis, we have considered only one root of
Eq.~(\ref{quadratic_for_z}), namely $\zeta_{e}=\zeta_{e}^{(1)}$. The reason is
that only this root provides an accurate transition to the well-established
SSI solution. The other root, $\zeta_{e}=\zeta_{e}^{(2)}$ has no SSI analog.
Besides, the corresponding value of $\zeta_{i}=\zeta_{e}^{(2)}+G_{i}$ becomes
fairly small and inefficient for driving the instabilities (see below).

To conclude this section, we note that in the long-wavelength limit, the
highest-order approximation to the reduced dispersion relation
(\ref{D_reduced}) describes the linear wave phase velocity relation%
\begin{equation}
\omega_{r}\approx\vec{k}\cdot\vec{V}_{0}+\omega_{De}(\vec{k})=\left[
\cos\theta+\zeta_{e}(\theta)\right]  kV_{0}. \label{omega_r=}%
\end{equation}
where $\zeta_{e}$ is the proper solution of Eq.~(\ref{Equation_for_zeta_e})
discussed above. In the LWL, this relation is common for all stable or
unstable waves, whatever the specific mechanism of wave generation. Notice the
linear $k$-scaling of the real wave frequency (and hence of all
Doppler-shifted frequencies, $\omega_{Ds}$). The next-order approximation
provides the instability growth/damping rates, which are different for
different physical mechanisms. The corresponding analysis will be performed in
the following section.

\subsection{First-order approximation: instability growth/damping rates.
Different physical
mechanisms\label{Instability growth rates: physical mechanisms}}

To determine specific mechanisms of instability generation, we need to
consider the next, i.e., first-order, approximation with respect to the small
parameters $\left\vert A_{e,j}\right\vert ,\left\vert B_{e,j}\right\vert
,k_{\parallel}^{2}/k_{\perp}^{2},k^{2}\lambda_{D}^{2}$ introduced by
Eq.~(\ref{small}). To find the instability growth/damping rates, $|\gamma
|\ll\omega_{r}$, according to Eq.~(\ref{gamma_small}), we need to linearize
the RHS of Eq.~(\ref{D_reduced}) with respect to the above small parameters
and retain only the imaginary part of $D(\omega,\vec{k})$. (The real part of
the first-order term in the Taylor expansion of $D(\omega,\vec{k})$ will
provide just a small correction to the wave phase velocity relation and will
be of no interest to us.) Given the known solution for $\omega_{De}(\vec
{k})=\zeta_{e}(\theta)kV_{0}$, and hence for all $\omega_{Dj}(\vec{k}%
)=\omega_{De}+\vec{k}\cdot\vec{U}_{j}=[\zeta_{e}(\theta)+G_{j}(\theta)]kV_{0}%
$, finding the growth/damping rates becomes a straightforward procedure.

We start by calculating the denominator in the RHS of Eq.~(\ref{gamma_small}).
According to Eq.~(\ref{D_0}) and (\ref{ss_phase_velocity}), where $\omega
_{De}$ and all $\omega_{Dj}$ are known functions of $\omega\approx\omega_{r}$
determined to the leading (zeroth-order) accuracy (see above), we obtain:%
\begin{equation}
\frac{\partial D_{0}(\omega_{r},\vec{k})}{\partial\omega_{r}}=\sum_{j=1}%
^{p}\frac{\rho_{j}(\vec{k}\cdot\vec{U}_{j})}{(1+\kappa_{j}^{2})\omega_{Dj}%
^{2}\psi_{j}}. \label{d_Re(D)}%
\end{equation}

Calculating the numerator in the RHS of Eq.~(\ref{gamma_small}), i.e.,
$\operatorname{Im}D(\omega,\vec{k})$, is a more cumbersome procedure. In the
RHS\ of Eq.~(\ref{D_reduced}), the standalone terms $\propto A_{s}$, $B_{s}$
given by Eqs.~(\ref{A_s_opiat})--(\ref{ion_angles}) are small and can be used
to the leading-order accuracy, while the ratio $A_{j}/B_{j}$ requires a better
accuracy. Neglecting small terms $\propto i\omega_{De}/\nu_{en}$, but keeping
the first-order approximation with respect to $\left\vert \Omega
_{j}\right\vert /\nu_{jn}=|\omega-\vec{k}\cdot\vec{V}_{j}|/\nu_{jn}$, and bearing in mind that usually
$\nu_{en}\gg\nu_{jn}$, we obtain
\begin{align*}
\frac{\alpha_{j}A_{j}}{A_{e}}  &  \approx\frac{\omega_{De}\kappa_{e}^{2}%
\nu_{en}m_{e}(1-i\omega_{Dj}/\nu_{jn})}{\omega_{Dj}\nu_{jn}m_{i}(1+\kappa
_{e}^{2}k_{\parallel}^{2}/k_{\perp}^{2})[\left(  1-i\omega_{Dj}/\nu
_{jn}\right)  ^{2}+\kappa_{j}^{2}]}\\
&  \approx\frac{\omega_{De}}{\omega_{Dj}\psi_{j}(1+\kappa_{j}^{2})}\left(
1+i\ \frac{1-\kappa_{j}^{2}}{1+\kappa_{j}^{2}}\frac{\omega_{Dj}}{\nu_{jn}%
}\right)  ,
\end{align*}
so that%
\begin{equation}
\operatorname{Im}\left(  1+\sum_{j=1}^{p}\frac{\rho_{j}\alpha_{j}A_{j}}{A_{e}%
}\right)  \approx\sum_{j=1}^{p}\frac{(1-\kappa_{j}^{2})\rho_{j}\omega_{De}%
}{(1+\kappa_{j}^{2})^{2}\nu_{jn}\psi_{j}}. \label{imho}%
\end{equation}
Substituting Eq.~(\ref{imho}) into Eq.~(\ref{d_Re(D)}) and slightly
redistributing the terms in the RHS\ of Eq.~(\ref{gamma_small}), we obtain the
following interim expression for the instability growth rate:%
\begin{widetext}
\begin{align}
&  \gamma\approx-\ \frac{\omega_{De}}{\sum_{j=1}^{p}\left.  \rho_{j}(\vec
{k}\cdot\vec{U}_{j})\right/  [(1+\kappa_{j}^{2})\omega_{Dj}^{2}\psi_{j}]}%
\sum_{j=1}^{p}\frac{\rho_{j}}{\left(  1+\kappa_{j}^{2}\right)  \omega_{Dj}%
\psi_{j}}\nonumber\\
&  \times\left\{  \frac{\omega_{Dj}}{\nu_{j}}\left[
\overset{\mathrm{Farley-Buneman}}{\overbrace{\frac{1-\kappa_{j}^{2}}%
{1+\kappa_{j}^{2}}}}-\overset{\text{\textrm{Charge }}\mathrm{Separation}%
}{\overbrace{\frac{\left(  1+\kappa_{j}^{2}\right)  \nu_{jn}^{2}}{\omega
_{pj}^{2}}}}\right]  +~\overset{\mathrm{Diffusion}\text{ Losses}%
}{\overbrace{\operatorname{Im}\left(  A_{j}-A_{e}\right)  }}~\right.
\nonumber\\
&  +~\left.  \overset{\mathrm{Ion}\text{ }\mathrm{Thermal}%
}{\overbrace{\operatorname{Im}\frac{1}{\mu_{j}}\left(  \frac{2A_{j}}{3}%
+B_{j}\right)  }}\overset{\mathrm{Electron}\text{ }\mathrm{Thermal}%
}{~-~\overbrace{\operatorname{Im}\frac{1}{\mu_{e}}\left(  \frac{2A_{e}}%
{3}+B_{e}\right)  }}\right\} \nonumber\\
&  =\gamma_{\mathrm{FB}}-\gamma_{\mathrm{CS}}+\gamma_{\mathrm{DL}}%
+\gamma_{\mathrm{IT}}+\gamma_{\mathrm{ET}}, \label{gamma_struct}%
\end{align}
\end{widetext}
where $
\omega_{pj}\equiv\left(  {e^{2}n_{e0}}/{\epsilon_{0}m_{j}}\right)  ^{1/2}$
is the plasma frequency of the $j$-th ion species. The labels over the
braces, along with the corresponding acronyms in the subscripts at the
bottom line of Eq.~(\ref{gamma_struct}), show the physical interpretation of
each term. They have a straightforward meaning. The
Farley-Buneman\ (\textquotedblleft FB\textquotedblright)\ instability term
originates from Eq.~(\ref{imho}). The label \textquotedblleft Charge
Separation\textquotedblright\ (\textquotedblleft CS\textquotedblright) means
a small deviation from quasi-neutrality; the corresponding term stems from
the $k^{2}\lambda_{D}^{2}/A_{e}$ term in the RHS\ of Eq.~(\ref{D_reduced}),
though without the corresponding multiplier in the square bracket (the terms
$\propto A_{e}$ and $B_{e}$ multiplied by $k^{2}\lambda_{De}^{2}/A_{e}$ would
lead to negligibly small, second-order corrections). The label
\textquotedblleft Diffusion Losses\textquotedblright\ (\textquotedblleft
DL\textquotedblright) denotes the diffusion losses caused by density gradients
formed within the given compression/rarefaction wave. Depending on the
parameters and wave characteristics, the \textquotedblleft
FB\textquotedblright, \textquotedblleft ET\textquotedblright, and
\textquotedblleft IT\textquotedblright\ mechanisms are responsible for driving
the FBI, ETI, and ITI, respectively, while the \textquotedblleft
DL\textquotedblright\ and \textquotedblleft CS\textquotedblright\ are
stabilizing\ (damping) mechanisms.

Before proceeding with the explicit expressions for the above terms, we
briefly discuss the physical mechanisms behind the wave damping and
instabilities. We start by discussing the wave damping mechanisms. The major
of the damping mechanisms, the diffusion losses of given particles of species
$s$ are caused by the ambipolar diffusion of the particles from the wave
density crests to the nearby wave troughs. This plasma particle diffusion is
caused by the wave spatial gradients of the regular particle pressure,
$\nabla(n_{s}T_{s})\propto i\vec{k}T_{s0}\delta n_{s}$ (assuming for
simplicity the isothermal regime). Within a given density wave, the particle
diffusion is always stabilizing. In the absence of instability excitation
mechanisms, the particle diffusion would eventually smear out any initially
created wave density perturbations, leading to the total wave disappearance.
The linear instability means that there should exist some physical mechanisms
that are able to reverse the stabilizing effect of the ambipolar diffusion and
lead to an exponential growth of the initial small wave perturbation. For a
physical explanation of the charge separation (CS) effect, see the appendix of
Ref.\ \onlinecite{Kovalev:Modeling08}.

Now we briefly discuss the instability driving mechanisms. The FBI is
driven by the ion inertia. In the wave frame of reference, this inertia,
through the $m_{s}(\vec{V}_{s}\cdot\nabla)\vec{V}_{s}$-term hidden within the
$m_{s}D_{s}\vec{V}_{s}/Dt$-term of Eq.~(\ref{my_fluid_equations_mom}), creates
an additional \textquotedblleft kinetic\textquotedblright\ pressure
perturbation, $m_{s}(\vec{V}_{s}\cdot\nabla)\vec{V}_{s}\rightarrow\nabla
(m_{s}V_{s}^{2}/2)\propto i m_{s}\vec{k}\cdot(\vec{V}_{s}-\vec{V}_{\mathrm{ph}})\delta
\vec{V}_{s}$, where $\vec{V}_{\mathrm{ph}}$ is the wave phase velocity. For
sufficiently strong driving electric field, $\vec{E}_{0}$, and properly
oriented (with respect to $\vec{E}_{0}$ and $\vec{B}_{0}$) wavevector $\vec
{k}$, this additional pressure may be in antiphase to the wave perturbation of
the regular plasma pressure $\propto T_{s0}\delta n_{s}$, overpower the
latter, and hence drive the linear instability.

For the two thermal-driven instabilities, ETI and ITI, the additional
pressure is created by wave modulations of the total ohmic heating described
by the first term in the RHS of Eq.~(\ref{my_fluid_equations_temperat}). The
modulated heating of plasma particles is caused by the wave electrostatic
field, $\delta\vec{E}$. Balanced by collisional cooling, this heating leads to
local modulations of the corresponding species temperature, $\delta T_{s}$.
Similarly to the FBI, for the properly oriented wavevector $\vec{k}$, the
additional pressure $\propto n_{s0}\delta T_{s}$ may reverse the sign of the
total wave pressure perturbation $\propto(T_{s0}\delta n_{s}+n_{s0}\delta
T_{s})$ and drive the instability.

The explicit expressions for the specific partial growth/damping rates,
calculated to the leading-order accuracy, are given by
\begin{widetext}
\begin{subequations}
\label{gamma_DL,IT,ET}%
\begin{align}
&  \gamma_{\mathrm{FB}}-\gamma_{\mathrm{CS}}\nonumber\\
&  =\left.  \sum_{j=1}^{p}\frac{\rho_{j}(-\omega_{De})}{(1+\kappa_{j}^{2}%
)\nu_{jn}\psi_{j}}\left[  \frac{1-\kappa_{j}^{2}}{1+\kappa_{j}^{2}}%
-\frac{(1+\kappa_{j}^{2})\nu_{jn}^{2}}{\omega_{pj}^{2}}\right]  \right/
\sum_{j=1}^{p}\frac{\rho_{j}(\vec{k}\cdot\vec{U}_{j})}{(1+\kappa_{j}%
^{2})\omega_{Dj}^{2}\psi_{j}},\label{gamma_FB-CS}\\
& \nonumber\\
&  \gamma_{\mathrm{DL}}=\left.  -\sum_{j=1}^{p}\frac{\rho_{j}k^{2}V_{Tj}^{2}%
}{(1+\kappa_{j}^{2})\omega_{Dj}\psi_{j}\nu_{jn}}\ \left[  \frac{T_{e0}\psi
_{j}}{T_{j0}}-\frac{\omega_{De}}{(1+\kappa_{j}^{2})\omega_{Dj}}\right]
\right/  \sum_{j=1}^{p}\frac{\rho_{j}(\vec{k}\cdot\vec{U}_{j})}{(1+\kappa
_{j}^{2})\omega_{Dj}^{2}\psi_{j}},\label{gamma_DL}\\
& \nonumber\\
&  \gamma_{\mathrm{IT}}=\nonumber\\
&  =\left.  \sum_{j=1}^{p}\frac{4m_{n}}{3(m_{j}+m_{n})}\ \frac{\rho
_{j}(-\omega_{De})\kappa_{j}kU_{j}(\kappa_{j}\cos\chi_{j}-\sin\chi_{j}%
)}{(1+\kappa_{j}^{2})^{2}\omega_{Dj}\psi_{j}\delta_{jn}\nu_{jn}}\right/
\sum_{j=1}^{p}\frac{\rho_{j}(\vec{k}\cdot\vec{U}_{j})}{(1+\kappa_{j}%
^{2})\omega_{Dj}^{2}\psi_{j}},\label{gamma_IT}\\
& \nonumber\\
&  \gamma_{\mathrm{ET}}=\frac{4kV_{0}\delta_{en}\nu_{en}\sin\theta}%
{3(\omega_{De}^{2}+\delta_{en}^{2}\nu_{en}^{2})\kappa_{e}}\nonumber\\
&  \times\left.  \sum_{j=1}^{p}\frac{\rho_{j}\omega_{De}}{(1+\kappa_{j}%
^{2})\omega_{Dj}\psi_{j}}\left(  1-\frac{T_{e0}kV_{Tj}^{2}\omega_{De}\psi
_{j}\kappa_{e}}{2T_{j0}\nu_{jn}V_{0}\delta_{en}\nu_{en}\sin\theta}\right)
\right/  \sum_{j=1}^{p}\frac{\rho_{j}(\vec{k}\cdot\vec{U}_{j})}{(1+\kappa
_{j}^{2})\omega_{Dj}^{2}\psi_{j}}, \label{gamma_ET}%
\end{align}
\end{subequations}
\end{widetext}
where the angles $\chi_{j}(\theta)$ are defined by Eq.~(\ref{ion_angles}). As
discussed in Sec.~\ref{Wave phase velocity relation}, for any allowed
linear-wave modes, $\omega_{De}$ is always negative, while all corresponding
$\omega_{Dj}=\omega_{De}+\vec{k}\cdot\vec{U}_{j}$ are positive. The
diffusion loss rate, $\gamma_{\mathrm{DL}}$, is always negative,
whereas in order to drive the FBI ($\gamma_{\mathrm{FB}}-\gamma_{\mathrm{CS}%
}>0$) the square bracket in the RHS\ of Eq.~(\ref{gamma_FB-CS}) has to be
positive.

In Eq.~(\ref{gamma_FB-CS}), we have combined the Farley-Buneman driving
mechanism ($\gamma_{\mathrm{FB}}$, see the first term in the square brackets)
with the charge-separation losses ($\gamma_{\mathrm{CS}}$, see the second term
in the square brackets) in order to emphasize the possible detrimental effect
of small deviations from quasi-neutrality on the FBI
\cite{Rosenberg:Farley98}. In the Earth's ionosphere, due to a sufficiently
high plasma density, the CS effect is usually negligible ($\nu_{jn}^{2}%
\ll\omega_{pj}^{2}$), although it always should be taken into account in PIC
simulations \cite{Oppenheim:Ion04}. In the solar chromosphere, we cannot
exclude the efficiency of the CS effect in some regions. For a sufficiently
low plasma density leading to $\nu_{jn}^{2}>\omega_{pj}^{2}$, the FBI cannot
be excited regardless of the imposed electric-field strength. The finite ion
magnetization, $\kappa_{j}^{2}\gtrsim1$, only aggravates the situation,
especially for $\kappa_{j}^{2}>1$, when even the FBI mechanism itself becomes stabilizing
\cite{Dimant:Ion04}. For other instabilities, the ITI and ETI, the CS
effect increases the instability threshold, but it is not totally detrimental,
regardless of the ratio $\nu_{jn}/\omega_{pj}$.

Being interested mostly in the minimal instability threshold, we can simplify
our treatment further by extending the assumed LWL to even longer wavelengths
that obey stronger conditions:%
\begin{equation}
kU_{j},|\omega_{Ds}|\ll\delta_{en}\nu_{en}. \label{SWL}%
\end{equation}
Usually $\delta_{en}\nu_{en}\ll\nu_{jn}$, so that the wavelengths obeying
these conditions are typically much longer than those defining the LWL,
see~Eq.~(\ref{long_conditions}). We will name the new limit imposed by
Eq.~(\ref{SWL}) the superlong-wavelength\ limit (SLWL). In accord with the
SLWL conditions, we can neglect in Eq.~(\ref{gamma_ET}) $\omega_{De}^{2}$
compared to $\delta_{en}^{2}\nu_{en}^{2}$, as well as the second term in the
first-summation parentheses compared to $1$. This will minimize the
threshold-field value along any given $\vec{k}$-direction (i.e., for given
$\theta$). According to zeroth-order Eq.~(\ref{D_0}), the remaining summation
in the numerator of Eq.~(\ref{gamma_ET}) equals $-1$, so that in the SLWL
$\gamma_{\mathrm{ET}}$ reduces to a much simpler expression,%
\begin{equation}
\gamma_{\mathrm{ET}}\approx -\ \left.  \frac{4kV_{0}\sin\theta}{3\delta_{en}%
\nu_{en}\kappa_{e}}\right/  \sum_{j=1}^{p}\frac{\rho_{j}(\vec{k}\cdot\vec
{U}_{j})}{(1+\kappa_{j}^{2})\omega_{Dj}^{2}\psi_{j}}.
\label{gamma_ET_long_wavelength}%
\end{equation}

Now we check the SSI case, $p=1$ ($j\rightarrow i$). In that case,
Eq.~(\ref{gamma_DL}) rate reduces to%
\begin{equation}
\gamma_{\mathrm{DL}}=-\ \frac{\omega_{Di}k^{2}V_{Ti}^{2}}{(\vec{k}\cdot\vec
{U}_{i})\nu_{in}}\left[  \frac{T_{e}\psi_{i}}{T_{i}}-\frac{\omega_{De}%
}{(1+\kappa_{i}^{2})\omega_{Di}}\right]  . \label{gamma_diff_ss}%
\end{equation}
Using the expressions for $\omega_{De,i}$ from Eq.~(\ref{ss_phase_velocity})
and combining Eq.~(\ref{gamma_diff_ss}) with similarly calculated
$\gamma_{\mathrm{FB}}$ and $\gamma_{\mathrm{CS}}$ we obtain the SSI expression
for the combined growth/damping rate which includes no thermal driving:
\begin{align}
&  \gamma_{\mathrm{FB}}-\gamma_{\mathrm{CS}}+\gamma_{\mathrm{DL}}\nonumber\\
& \! =\frac{\psi_{i}\omega_{Di}^{2}}{[1+(1+\kappa_{i}^{2})\psi_{i}]\nu_{in}%
}\left[  1-\kappa_{i}^{2}-\frac{k_{\perp}^{2}C_{s}^{2}}{\omega_{Di}^{2}}%
-\frac{(1+\kappa_{i}^{2})^{2}\nu_{in}^{2}}{\omega_{pi}^{2}}\right]\!\!  ,
\label{gamma_FB-CS+DL_ss}%
\end{align}
where $\omega_{Di}=\vec{k}\cdot\vec{U}_{i}/[1+\left(  1+\kappa_{i}^{2}\right)
\psi_{i}]$, while $C_{s}$ is the isothermal ion-acoustic speed defined by
Eq.~(\ref{V_0_approx}). Equation~(\ref{gamma_FB-CS+DL_ss}) agrees with the
previous results for the arbitrary ion magnetization, see, e.g., Eq.~(6) from
Ref.\ \onlinecite{Dimant:Model03}, except for the last term in the square brackets which
generalizes the CS term from Ref.\ \onlinecite{Rosenberg:Farley98} to $\kappa_{i}^{2}\sim1$.

Now we note that in the SLWL all driving/damping rates $\gamma_{\mathrm{s}}$,
except $\gamma_{\mathrm{ET}}$ (see below), have a simple quadratic
$k$-scaling: $\gamma_{\mathrm{s}}\propto k^{2}$. To establish this, it is
sufficient to assume the linear $k$-dependence of $\omega_{r}\propto k$. This
is clear from $\omega_{Ds}\propto k$, in full consistency with Eq.~(\ref{D_0})
and its solutions (discussed in Sec.~\ref{Wave phase velocity relation}).
Setting $\omega_{Ds}\propto k$ in Eq.~(\ref{gamma_DL,IT,ET}) with
Eq.~(\ref{gamma_ET}) replaced by Eq.~(\ref{gamma_ET_long_wavelength}), one can
easily establish the $\gamma_{s}\propto k^{2}$ scaling. This common scaling
for all $\gamma_{\mathrm{s}}=0$ automatically makes the threshold
field along the given $\vec{k}$-direction to be $k$-independent -- the
well-established fact for the pure FBI in the LWL fluid-model approximation,
see, e.g., Refs.\ \onlinecite{Fejer:Theory84,Farley:Theory85}. If the FBI is the dominant
instability driver, as in most of the E-region ionosphere, then
within the entire LWL the growth rate $\gamma\propto k^{2}$, so that its
maximum is usually reached beyond the LWL (see also Appendix~\ref{APPENDIX A}).

If the dominant instability driver is the ETI, as we observed in our recent
PIC\ simulations for some solar chromosphere parameters
\cite{Oppenheim:Newly20}, then the growth rate maximum is reached within the
LWL due to the competition between the two terms within the parentheses under
the first summation in Eq.~(\ref{gamma_ET}). In the SSI case of pure ETI
driving, we have%
\begin{widetext}
\begin{equation}
\gamma_{\mathrm{ET}}\approx-\ \frac{(1+\kappa_{i}^{2})\psi_{i}(\vec{k}%
\cdot\vec{U}_{i})}{[1+(1+\kappa_{j}^{2})\psi_{i}]^{2}}\frac{4kV_{0}\delta
_{en}\nu_{en}\sin\theta}{3(\omega_{De}^{2}+\delta_{en}^{2}\nu_{en}^{2}%
)\kappa_{e}}\left(  1-\frac{T_{e}k\omega_{De}V_{Ti}^{2}\psi_{i}\kappa_{e}%
}{2T_{i}\nu_{in}V_{0}\delta_{en}\nu_{en}\sin\theta}\right)  .
\label{gamma_ET_SS}%
\end{equation}
\end{widetext}
The first term in parentheses (i.e., $1$) reflects the local heating-cooling
balance, which is the crucial factor for the ETI. The second term $\propto
k\omega_{De}\propto k^{2}$ is responsible for the nonlocal temperature spread
within the wavelength due to the heat advection. Since $\omega_{De}$ is
negative (see Sec.~\ref{Wave phase velocity relation}), total $\gamma
_{\mathrm{ET}}$ can be positive for some $\vec{k}$ within the negative sector,
while for $\vec{k}$ within the positive sector of $\theta$, the rate
$\gamma_{\mathrm{ET}}$ is always negative, regardless of the $E_{0}$ value. In
the SLWL of $kU_{j}$, $\omega_{Ds}\ll\delta_{en}\nu_{en}$, neglecting
$\kappa_{i}^{2}\psi_{i}$, and taking $\vec{U}_{i}\approx\vec{V}_{0}$ (assuming
also $\kappa_{i}^{2}\ll1$), we obtain a much simpler relation:%
\begin{equation}
\gamma_{\mathrm{ET}}\simeq-\ \frac{4\psi_{i}k^{2}V_{0}^{2}\sin\theta\cos
\theta}{3\left(  1+\psi_{i}\right)  ^{2}\kappa_{e}\delta_{en}\nu_{en}}.
\label{gamma_ET_reduced}%
\end{equation}
For $m_{n}=m_{i}$, $\delta_{i}=1$, $(1+\kappa_{i}^{2})\psi_{i}\rightarrow
\psi_{i}$, and bearing in mind that $\kappa_{i}^{2}\psi_{i}=\kappa_{i}%
/\kappa_{e}\ll1$, Eq.~(\ref{gamma_IT_SS}) agrees with Eq.~(30) from
Ref.\ \onlinecite{Dimant:Ion04}. To the accuracy of the factor of order unity, this agrees with
the previous results, see, e.g., Eq.~(38) from Ref.\ \onlinecite{Dimant:Ion04}, neglecting
the term $\propto S^{2}$ originated there from the electron-temperature
dependence of $\nu_{en}$. Recall that, assuming elastic $e$-$n$ collisions
determined mostly by the electron polarization of the colliding neutral
particle, in this paper we ignore any temperature dependence of $\nu_{en}$. We
note that ignoring the $\propto S^{2}$ term leads to the absence of the
additional destabilizing ETI mechanism, which is, unlike that in
Eq.~(\ref{gamma_ET_reduced}), symmetric with respect to the sign of $\theta$, see Refs.\
\onlinecite{Dimant:Kinetic95b,Dimant:Physical97}.

Finally, we check the SSI case for the ion thermal driving. In the SSI case,
Eq.~(\ref{gamma_IT}) reduces to%
\begin{equation}
\gamma_{\mathrm{IT}}\approx\frac{4\psi_{i}k^{2}U_{i}^{2}m_{n}(\kappa_{i}%
\cos\chi_{i})(\kappa_{i}\cos\chi_{i}-\sin\chi_{i})}{3[1+(1+\kappa_{i}^{2}%
)\psi_{i}]^{2}(m_{n}+m_{i})\delta_{in}\nu_{in}}, \label{gamma_IT_SS}%
\end{equation}
which also agrees with the previous results \cite{Dimant:Ion04}.

\subsection{Threshold electric field\label{Threshold electric field}}

The threshold electric field for the combined instability (the FBI, ETI, and
ITI) is determined by equating the total growth rate to zero,
\begin{equation}
\gamma\equiv\gamma_{\mathrm{FB}}-\gamma_{\mathrm{CS}}+\gamma_{\mathrm{DL}%
}+\gamma_{\mathrm{IT}}+\gamma_{\mathrm{ET}}=0. \label{gamma=0}%
\end{equation}
where all $\gamma_{\mathrm{S}}$ are given by Eq.~(\ref{gamma_DL,IT,ET}). For a
given wave mode determined by its wavevector $\vec{k}$, we have obtained above
the zeroth-order solution for the real negative electron Doppler-shifted
frequency $\omega_{De}\approx\omega_{Der}=\zeta_{e}kV_{0}$, see
Eq.~(\ref{Omega_e^+}) or its simplified versions given by
Eqs.~(\ref{Omega_e_Solar_unmagnetized})-(\ref{interpolation}). The parameters
in these solutions are expressed in terms of $\vec{k}\cdot\vec{U}_{j}%
=G_{j}kV_{0}$, where $G_{j}$ is defined in Eq.~(\ref{kU_via_G}), and
$kU_{j}=kV_{0}/(1+\kappa_{j}^{2})^{1/2}$, see Eq.~(\ref{UU}) and
(\ref{VVU_abs_values}), i.e., eventually, in terms of the driving-field
amplitude, $E_{0}=V_{0}B_{0}$ and the wavevector $\vec{k}$. Then the
quantities $\omega_{Dj}=\left(  \zeta_{e}+G_{j}\right)  kV_{0}$, involved in
all $\gamma_{\mathrm{S}}$, become also functions of $E_{0}$. Given $\vec{k}$ and
the proper solution for $\zeta_{e}$, by solving Eq.~(\ref{gamma=0}) we obtain
the instability threshold $E_{0}=E_{\mathrm{Thr}}$. Bearing in mind the minimal threshold fields,
we will restrict our further treatment of wavelengths to the SLWL, in which
the scaling $\gamma\propto k^{2}$ holds for all instability
driving and loss mechanisms. This will allow us to cancel all $k$-related
factors and obtain the general, $k$-independent, minimum value of the
threshold field. While the $k$-dependence of $E_{\mathrm{Thr}}$ disappears,
the dependence on the $\vec{k}$ angles still holds and is crucial. Note that
total absence of real positive roots for $E_{\mathrm{Thr}}$ within a given
parameter domain means the linearly stable regime, regardless of the strength
of the imposed electric field $\vec{E}_{0}$.

To apply Eq.~(\ref{gamma=0}), we express $\omega_{De,j}$, $\vec{k}\cdot\vec
{U}_{j}$ and $U_{j}=V_{0}/(1+\kappa_{j}^{2})^{1/2}$ in terms of $\zeta_{e}$
and $V_{0}$. Leaving out in Eq.~(\ref{gamma_DL,IT,ET}) the inconsequential
common denominator $\sum_{j=1}^{p}\rho_{j}(\vec{k}\cdot\vec{U}_{j}%
)/[(1+\kappa_{j}^{2})\omega_{Dj}^{2}\psi_{j}]$, along with the remaining
$k$-factor, we obtain
\begin{widetext}
\begin{subequations}
\label{gamma_propto}%
\begin{align}
&  \gamma_{\mathrm{FB}}-\gamma_{\mathrm{CS}}\propto-V_{0}\zeta_{e}\sum
_{j=1}^{p}\frac{\rho_{j}}{(1+\kappa_{j}^{2})\psi_{j}\nu_{jn}}\left[
\frac{1-\kappa_{j}^{2}}{1+\kappa_{j}^{2}}-\frac{(1+\kappa_{j}^{2})\nu_{jn}%
^{2}}{\omega_{pj}^{2}}\right]  ,\label{gamma_FB_propto}\\
&  \gamma_{\mathrm{DL}}\propto-\sum_{j=1}^{p}\frac{\rho_{j}V_{Tj}^{2}%
}{(1+\kappa_{j}^{2})\zeta_{j}V_{0}\psi_{j}\nu_{jn}}\left[  \frac{T_{e}\psi
_{j}}{T_{j}}-\frac{\zeta_{e}}{(1+\kappa_{j}^{2})\zeta_{j}}\right]
,\label{gamma_DL_propto}\\
&  \gamma_{\mathrm{IT}}\propto-V_{0}\zeta_{e}\sum_{j=1}^{p}\frac{4m_{n}%
}{3(m_{j}+m_{n})}\ \frac{\rho_{j}(\kappa_{j}\cos\chi_{j}-\sin\chi_{j}%
)\kappa_{j}}{(1+\kappa_{j}^{2})^{5/2}\zeta_{j}\delta_{jn}\psi_{j}\nu_{jn}%
},\label{gamma_IT_propto}\\
&  \gamma_{\mathrm{ET}}\propto-\ \frac{4V_{0}\sin\theta}{3\delta_{en}\nu
_{en}\kappa_{e}}. \label{gamma_ET_propto}%
\end{align}
\end{subequations}
\end{widetext}
Here $\zeta_{j}=\zeta_{e}+G_{j}$, $G_{i}=(\cos\chi_{j})/(1+\kappa_{j}%
^{2})^{1/2}$, and the symbol ``$\propto$''
has a stronger meaning that just ``proportionality''; it implies a dropped common factor
for all $\gamma_{\mathrm{s}}$. Given the proper solution of Eq.~(\ref{gamma=0})
for the negative variable $\zeta_{e}$, as
discussed in Sec.~\ref{Wave phase velocity relation}, we obtain the general
expression for the total instability threshold field in the SLWL:%
\begin{widetext}
\begin{equation}
V_{\mathrm{Thr}}=\frac{E_{\mathrm{Thr}}}{B_{0}}=\left\{  \left.  \sum
_{j=1}^{p}\frac{\rho_{j}V_{Tj}^{2}}{(1+\kappa_{j}^{2})\psi_{j}\nu_{jn}%
\zeta_{j}}\left[  \frac{T_{e}\psi_{j}}{T_{j}}-\frac{\zeta_{e}}{(1+\kappa
_{j}^{2})\zeta_{j}}\right]  \right/  R\right\}  ^{1/2}, \label{V_Thr_SLWL}%
\end{equation}
where%
\begin{equation}
R \equiv(-\zeta_{e})\sum_{j=1}^{p}\frac{\rho_{j}}{(1+\kappa_{j}^{2}%
)\psi_{j}\nu_{jn}}\left[  \frac{1-\kappa_{j}^{2}}{1+\kappa_{j}^{2}}%
-\frac{(1+\kappa_{j}^{2})\nu_{jn}^{2}}{\omega_{pj}^{2}}
\frac{4(\kappa_{j}\cos\chi_{j}-\sin\chi_{j})m_{n}\kappa_{j}%
}{3(m_{j}+m_{n})(1+\kappa_{j}^{2})^{3/2}\delta_{jn}\zeta_{j}}\right]
-\frac{4\sin\theta}{3\delta_{en}\nu_{en}\kappa_{e}}. \label{RR}%
\end{equation}
\end{widetext}

We imply here only positive values of $R$. If some wave and plasma parameters
lead to $R<0$ then $V_{\mathrm{Thr}}$ becomes imaginary. As mentioned above,
this means that this group of parameters corresponds to a totally stable
situation, regardless of how strong is the driving electric field. The SLWL
solution for $V_{\mathrm{Thr}}$ provides the absolute combined-instability threshold
minimum for the entire range of $k$. In the general multi-species ion case,
however, it is usually hard to find explicit analytical expressions for the
optimal $\vec{k}$-direction. For a given set of parameters, the optimal angle
can be found numerically.

Below we discuss two particular cases that provide significant simplifications:
(1) single-species ions and (2) multi-species, but fully unmagnetized ions.

\subsubsection{Single-species ions}

In the SSI case, $p=1$, $\rho_{j}=1$, $j\rightarrow
i$, using the relation $\zeta_{i}=\zeta_{e}+(\cos\chi_{i})/(1+\kappa_{i}%
^{2})^{1/2}$ (see above) and Eq.~(\ref{ion_angles}), we obtain
\begin{align*}
&\zeta_{e} = -\ \frac{\left(  \cos\theta-\kappa_{i}\sin\theta\right)  \psi_{i}%
}{1+\left(1+\kappa_{i}^{2}\right)  \psi_{i}},\\
&\zeta_{i} = \frac
{\cos\theta-\kappa_{i}\sin\theta}{\left(  1+\kappa_{i}^{2}\right)  \left[
1+\left(  1+\kappa_{i}^{2}\right)  \psi_{i}\right]  }.
\end{align*}
Then the SSI threshold field reduces to%
\begin{align}
 & V_{\mathrm{Thr}} =\frac{E_{\mathrm{Thr}}}{B_{0}}=\frac{\left[  1+\left(
1+\kappa_{i}^{2}\right)  \psi_{i}\right]  \left(  1+\kappa_{i}^{2}\right)
^{1/2}C_{s}}{\left(  \cos\theta-\kappa_{i}\sin\theta\right)  P}%
,\nonumber\\
 & P \equiv\left[  \frac{1-\kappa_{i}^{2}}{1+\kappa_{i}^{2}}-\frac
{(1+\kappa_{i}^{2})\nu_{in}^{2}}{\omega_{pi}^{2}}-\frac{4m_{n}\kappa_{i}%
\sin\theta}{3(m_{i}+m_{n})(1+\kappa_{i}^{2})\delta_{in}\zeta_{i}}\right.
\nonumber\\
&  \left.  -~\frac{4(1+\kappa_{i}^{2})\left[  1+\left(  1+\kappa_{i}%
^{2}\right)  \psi_{i}\right]  \nu_{in}\sin\theta}{3\delta_{en}\nu_{en}%
\kappa_{e}\left(  \cos\theta-\kappa_{i}\sin\theta\right)  }\right]
^{1/2}, \label{V_SS_Thr}
\end{align}
where $C_{s}=[(T_{e}+T_{i})/m_{i}]^{1/2}$ is the conventional isothermal
ion-acoustic velocity (already invoked in Sec.~\ref{Ohmic heating}).

\subsubsection{Unmagnetized ions}

For unmagnetized, but multi-species, ions, $\kappa_{j}\ll1$, we have equal
$G_{j}\approx\cos\theta$ for all ion species. According to
Eqs.~(\ref{Omegas_simple}) and (\ref{Psi}), in the limit of totally neglected
ion magnetization, $\kappa_{j}=0$, all $p$ roots of linear Eq.~(\ref{D_0}) for
$\zeta_{e}$ degenerate into a single root with all $\zeta_{j}$ equal to the
same common value $\zeta_{i}=\zeta_{e}+\cos\theta$,
\begin{equation}
\zeta_{e}=-\ \frac{\Psi\cos\theta}{1+\Psi},\qquad\zeta_{i}=\frac{\cos\theta
}{1+\Psi}. \label{Omega_e_unmagnetized}%
\end{equation}
Furthermore, for $\kappa_{j}\ll1$ the ITI driving term, $\gamma_{\mathrm{IT}}%
$, is small and can be neglected. As a result, after additionally canceling
the common factor $k$, Eqs.~(\ref{gamma_FB_propto})--(\ref{gamma_ET_propto})
reduce to much simpler relations:
\begin{subequations}
\label{gamma_propto_unmagnet}%
\begin{align}
&  \gamma_{\mathrm{FB}}-\gamma_{\mathrm{CS}}\propto\frac{\Psi V_{0}\cos\theta
}{1+\Psi}\sum_{j=1}^{p}\frac{\rho_{j}}{\psi_{j}\nu_{jn}}\left(  1-\frac
{\nu_{jn}^{2}}{\omega_{pj}^{2}}\right)  ,\label{gamma_FB_propto_unmagnet}\\
&  \gamma_{\mathrm{DL}}\propto-\ \frac{1+\Psi}{V_{0}\cos\theta}\sum_{j=1}%
^{p}\frac{\rho_{j}V_{Tj}^{2}}{\nu_{jn}}\left(  \frac{T_{e}}{T_{j}}+\frac{\Psi
}{\psi_{j}}\right)  ,\label{gamma_DL_propto_unmagnet}\\
&  \gamma_{\mathrm{ET}}\propto-\ \frac{4V_{0}\sin\theta}{3\delta_{en}\nu
_{en}\kappa_{e}}. \label{gamma_ET_propto_unmagnet}%
\end{align}
Introducing temporary notations%
\end{subequations}
\begin{align}
K  &  =\sum_{j=1}^{p}\frac{\rho_{j}V_{Tj}^{2}}{\nu_{jn}}\left(  \frac{T_{e}%
}{T_{j}}+\frac{\Psi}{\psi_{j}}\right)  ,\nonumber\\
M  &  =\Psi\sum_{j=1}^{p}\frac{\rho_{j}}{\psi_{j}\nu_{jn}}\left(  1-\frac
{\nu_{jn}^{2}}{\omega_{pj}^{2}}\right)  ,\qquad N=\frac{4\left(
1+\Psi\right)  }{3\delta_{en}\nu_{en}\kappa_{e}}, \label{MNK}%
\end{align}
we write the instability threshold for unmagnetized ions as%
\begin{align}
V_{\mathrm{Thr}}  &  =\frac{E_{\mathrm{Thr}}}{B_{0}}=\frac{1+\Psi}{\cos\theta
}\left(  \frac{K}{M-N\tan\theta}\right)  ^{1/2}\nonumber\\
&  =\left(  1+\Psi\right)  \left[  \frac{2K}{M\left(  1+\cos2\theta\right)
-N\sin2\theta}\right]  ^{1/2} \label{V_Thr_unmagnetized}%
\end{align}
Here, the term $\propto M$ stems from the FBI driving (combined with the
charge-separation damping $\propto\nu_{jn}^{2}/\omega_{pj}^{2}$), while the
term $\propto N$ stems from the ETI driving.
Equation~(\ref{V_Thr_unmagnetized}) keeps virtually the same flow-angle
restrictions for the instability as does the simpler SSI model
\cite{Fejer:Theory84,Dimant:Physical97,Dimant:Ion04}. In particular, for the
pure FBI the cone of allowed angles $\theta$ is symmetric around the $\vec
{E}_{0}\times\vec{B}_{0}$-drift direction $\theta=0^{\circ}$, while for the
pure ETI the allowed cone is situated around the negative bisector of
$\theta=-45^{\circ}$. At the positive domain of $\theta$, the ETI mechanism
becomes stabilizing (as does the FBI mechanism for $\nu_{jn}>\omega_{pj}$),
regardless of the electric-field strength.

The case of unmagnetized ions allows one to explicitly obtain the optimal
angles of $\vec{k}$ corresponding to the minimum values of $V_{\mathrm{Thr}}$
(or $E_{\mathrm{Thr}}$). In the main semi-quadrant of $\theta$, where\ $\cos
\theta\geq0$, the optimum angle $\theta_{\mathrm{opt}}$ is unambiguously
determined by%
\begin{equation}
\theta_{\mathrm{opt}}=-\ \frac{1}{2}\arctan\frac{N}{M}, \label{theta_opt}%
\end{equation}
with the corresponding minimum threshold values given by%
\begin{equation}
\left(  V_{\mathrm{Thr}}\right)  _{\min}=\frac{\left(  E_{\mathrm{Thr}%
}\right)  _{\min}}{B_{0}}=2\left(  1+\Psi\right)  \sqrt{  \frac{K}%
{M+\sqrt{M^{2}+N^{2}}}}. \label{V_Thr_min}%
\end{equation}
As might be expected, in the limiting cases of $N=0$ (the pure FBI) or $M=0$
(the pure ETI) the optimal angles reduce to $\theta_{\mathrm{opt}}=0^{\circ}$
or $\theta_{\mathrm{opt}}=-45^{\circ}$, respectively. The SLWL instability
threshold values given by Eq.~(\ref{V_Thr_min}) represent the global minimum
of the combined instability threshold for the unmagnetized multi-species ions
in the entire range of $\vec{k}$.

\section{ARBITRARY WAVELENGTHS\label{Instability dispersion relation for arbitrary wavelengths}}

In this section, we briefly discuss the general dispersion relation for
arbitrary wavelengths and give examples of its numeric solution.

First, we summarize the general multi-fluid model dispersion for arbitrarily
magnetized particles, see Eqs.~(\ref{disperga_snova}), (\ref{A_s_opiat}%
)--(\ref{ion_angles}). It can be re-written in a more compact way as%
\begin{equation}
1+\sum_{j=1}^{p}\frac{\lambda_{Dj}^{2}F_{j}}{\lambda_{De}^{2}F_{e}}%
=k^{2}\lambda_{De}^{2}F_{e},\label{General_dispersion_compact}%
\end{equation}
where
\begin{subequations}
\label{F,A,B_s_general}%
\begin{align}
F_{s} &  = A_{s}\left[  1-\left(  1+\frac{2}{3\mu_{s}}\right)  A_{s}%
-\frac{B_{s}}{\mu_{s}}\right]  ^{-1},\label{F_s_general}\\
A_{s} &  =-i\ \frac{V_{Ts}^{2}}{\nu_{sn}\omega_{Ds}}\left(  \frac
{W_{s}k_{\perp}^{2}}{W_{s}^{2}+\kappa_{s}^{2}}%
+\frac{k_{\parallel}^{2}}{W_{s}}\right)  ,\label{A_s_general}\\
B_{s} &  =\frac{4m_{n}\left[  W_{s}(\vec{k}_{\perp}\cdot\vec{V}%
_{s0})-\kappa_{s}\vec{k}_{\perp}\cdot(\vec{V}_{s0}\times\hat{b})\right]
}{3\omega_{Ds}\left(  m_{n}+m_{s}\right)  \left(  W_{s}^{2}%
+\kappa_{s}^{2}\right)  }\ ,\label{B_s_general}%
\end{align}
\end{subequations}
\begin{subequations}
\label{Remaining_1,2,3}%
\begin{align}
W_{s} &  =1-\frac{i\omega_{Ds}}{\nu_{sn}},\;\;\;\;\; \omega_{Ds}%
=\omega-\vec{k}\cdot\vec{V}_{s0},\;\;\;\;\; \lambda_{Ds}^{2}=\frac{\epsilon
_{0}T_{s0}}{q_{s}^{2}n_{s0}},\label{Remaining_1}\\
\vec{V}_{s0} &  =\left.  \left(  \frac{q_{s}\vec{E}_{0}}{m_{s}\nu_{sn}}%
+\kappa_{s}^{2}\vec{V}_{0}\right)  \right/  \left(  1+\kappa_{s}^{2}\right)
,\;\;\;\;\;\; \kappa_{s}=\frac{q_{s}B_{0}}{m_{s}\nu_{sn}},\label{Remaining_2}\\
\mu_{s} &  =1+i\ \frac{2m_{s}\nu_{sn}}{(m_{s}+m_{n})\omega_{Ds}},\qquad\xi
_{s}\equiv1+\frac{i\delta_{sn}\nu_{sn}}{\omega_{Ds}}.\label{Remaining_3}%
\end{align}
and $\vec{E}_0$ is the $\vec{E}_0\times\vec{B}_0$-drift velocity.
Here, the subscript $j$ describes different ion species, $j=1,2,...p$, while
the more general subscript $s$ includes each ion species ($s=j$) and electrons ($s=e$).

All variables and parameters in Eq.\ (\ref{General_dispersion_compact}) are written in the neutral-component frame of reference. If the neutral flow, presumed locally uniform, shearless, and quasi-stationary, moves in a laboratory frame with the non-relativistic velocity $\vec{V}_n$, then the electric field in Eq.\ (\ref{Remaining_1,2,3}), in terms of the electric field in the laboratory frame, $\vec{E}'_0$, is given by $\vec{E}_0 \approx \vec{E}'_0 - \vec{V}_n\times \vec{B}_0$ ($|\vec{E}'_0|, E_0 \ll cB_0$). In the same laboratory frame, the Doppler shifted wave frequency, $\omega'$, is given by $\omega'\approx \omega + \vec{k}\cdot\vec{V}_n$.

Before presenting examples of the real wave frequency and growth rates found by numerically solving Eq.\ (\ref{General_dispersion_compact}), we discuss distinct signatures of the pure thermal instabilities versus the pure Farley-Buneman instability. Waves driven by the pure ETI has three distinct features: (1) for unmagnetized ions, the preferred wavevectors tend to group around the bi-sector between the directions of the $\vec{E}_0\times\vec{B}_0$-drift velocity and the $-\vec{E}_0$ direction, i.e., where the corresponding growth rate is maximized, while the preferred direction for the FBI-driven waves is along the $\vec{E}_0\times\vec{B}_0$-drift velocity, (2) the wave perturbations of the electron temperature are mostly in anti-phase to the wave perturbations of the plasma density, while for the FBI-driven waves the corresponding wave perturbations are mostly in phase, (3) the typical wavelengths of the ETI-driven waves are usually much longer than those of the FBI-driven waves\cite{Dimant:Physical97}. For the pure ITI-driven waves, feature (1) is more complicated than for the pure ETI because the ITI is mostly pronounced if ions are partially magnetized, feature (2) stays the same as for the ETI, while feature (3) does not hold for the ITI-driven waves (the typical wavelengths of these waves are comparable to the wavelengths of the FBI-driven waves \cite{Dimant:Ion04}). The phase shift between the temperature perturbations (feature 2) can be identified in simulations of the instability (such nonlinear simulations are beyond the scope of this paper), while the preferred wavevector directions and wavelengths can be traced directly from the predicted growth rates.

Figures\ \ref{Real_frequencies} and \ref{Growth_rates} show examples of the numerical solution of Eq.~(\ref{General_dispersion_compact}) for the real and imaginary parts of the wave frequency, respectively, $\omega$, using different values of the driving electric field. The other parameters used here correspond to those employed for our recent fluid-model solar chromosphere simulations using the fluid-model Ebysus code\cite{Evans:Multi02_arxiv}. The major parameters used in these calculation are listed in the Table 1 of Ref.~\onlinecite{Evans:Multi02_arxiv}. The minimum threshold field for the chosen parameters is about $E_{\mathrm{Thr}}\approx 4.4$~eV. These figures show that as long as the driving field is not very far above the $E_{\mathrm{Thr}}$ the ETI seems to be a dominant instability mechanism. This can be easily seen from the above signatures (1) and (3): the preferred $\vec{k}$-directions tend to the $-45^\circ$ bisector and waves tends to smaller $k$ (longer wavelengths). As the driving field increases, the entire unstable region expands with the maximum growth rate shifting to larger $k$ (shorter wavelengths), while the preferred $\vec{k}$-directions start deviating initially closer to the horizontal $\vec{E}_0\times\vec{B}_0$-direction (typical for the FBI-driven waves) and then rotating further up to the vertical $\vec{E}_0$-direction. The latter has no simple explanation.
\begin{figure}
\noindent\includegraphics[width=0.48\textwidth]{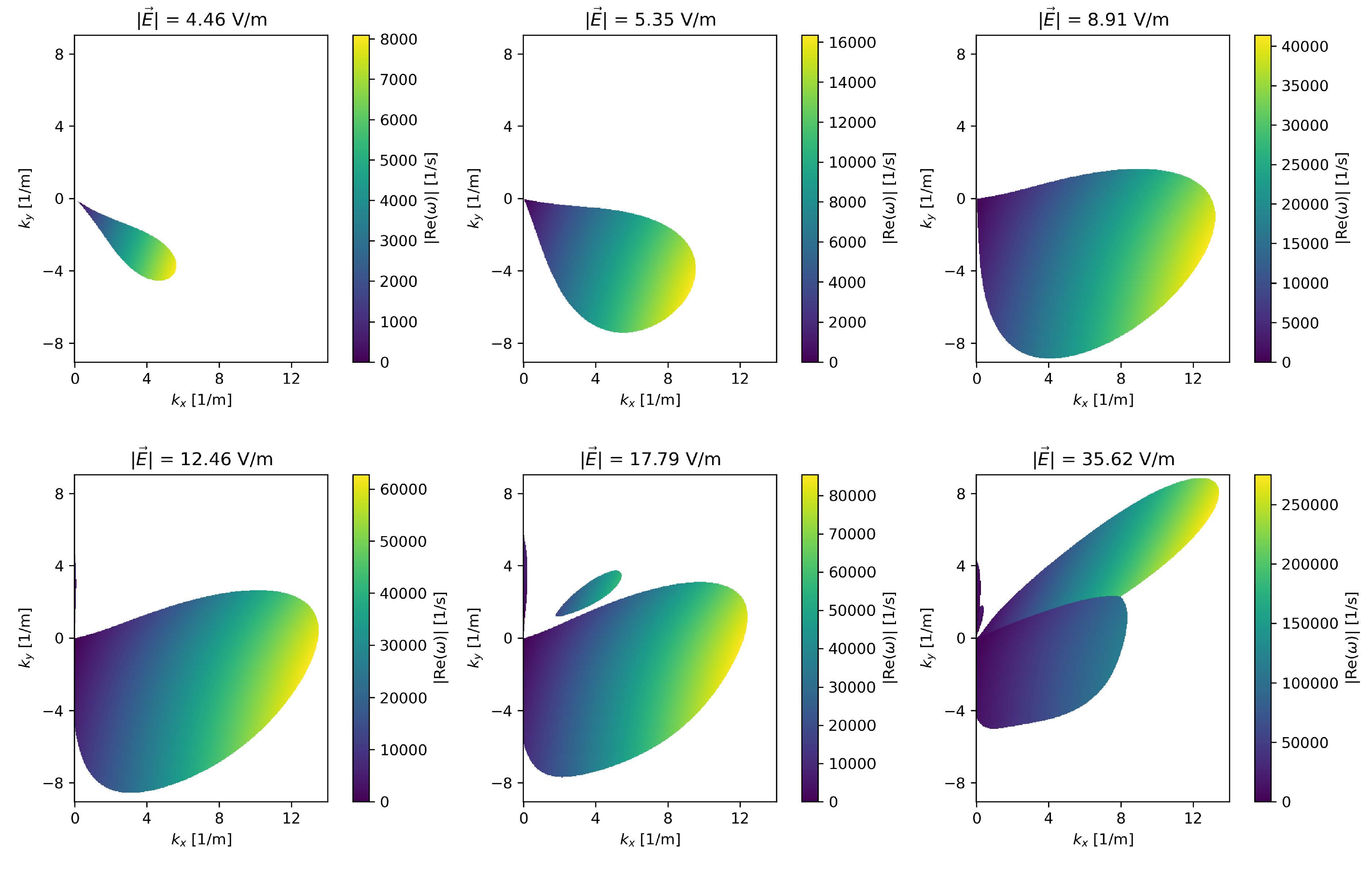}
\caption{\label{Real_frequencies}Examples of a numerical solution of Eq.~(\ref{General_dispersion_compact}) for the real part of the wave frequency, $\omega_r = \mathrm{Re}(\omega)$, for several values of the driving electric field $E_0 = |\vec{E}|$ shown on top of each plot. Only the areas where $\gamma > 0$ are shown. The driving electric field $\vec{E}_0$ is directed along the vertical $k_y$-axis, while the $\vec{E}_0\times\vec{B}_0$-drift velocity is directed along the horizontal $k_x$-axis.}
\end{figure}
\begin{figure}
\noindent\includegraphics[width=0.48\textwidth]{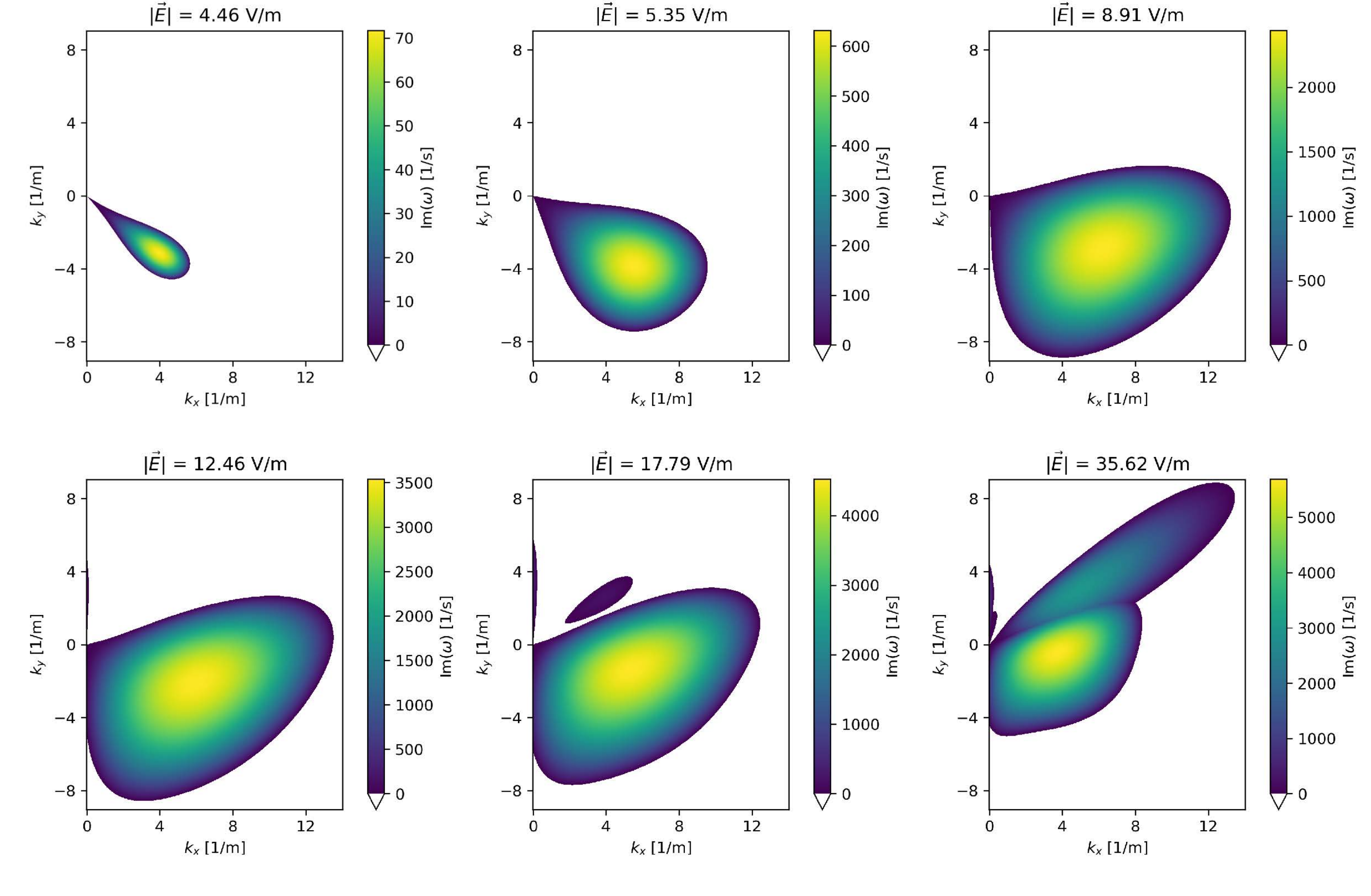}
\caption{\label{Growth_rates}Examples of a numerical solution of Eq.~(\ref{General_dispersion_compact}) for the imaginary part of the wave frequency, $\gamma = \mathrm{Im}(\omega)$, for the same values of the driving electric field as in Fig.\ \ref{Real_frequencies}. Only the areas where $\gamma > 0$ are shown.}
\end{figure}
At the driving field of $E_0 = 35.62$~V/m, which exceeds the minimum threshold field by an almost order of magnitude, we see two overlapping, but distinct, areas of short-wavelength unstable waves. It is possible, however, that this feature is a consequence of the restrictive fluid-model treatment. A more accurate kinetic approach may result in smearing these distinct areas. The main point, however, is that even our purely fluid-model treatment leads to a restricted area of linearly unstable waves in the $\vec{k}$-space (in full agreement with the analysis of Appendix~\ref{APPENDIX A}. This gives one a solid possibility to safely simulate $\vec{E}\times\vec{E}$ instabilities, using fluid-model codes without fear that such simulation may ``blow up'' at the short-wavelength band.

\section{SUMMARY AND CONCLUSIONS\label{SUMMARY AND CONCLUSIONS}}

This paper presents a theoretical analysis of a combined
Thermal-Farley-Buneman Instability (TFBI). This combined instability includes
the following components: the Farley-Buneman instability (FBI),
electron-thermal instability (ETI), and ion-thermal instability (ITI). All
these low-frequency, electrostatic, and inherently collisional plasma instabilities are
developed in weakly ionized, highly dissipative, and moderately magnetized
media, such as the solar chromosphere, lower Earth's ionosphere, the
corresponding regions of other star and planetary atmospheres, and potentially
in cometary tails, molecular clouds, accretion disks, etc. In this paper, we restrict our
analytic treatment to the linear theory of the TFBI. This theory is developed
in the framework of the 5-moment multi-fluid\ set of equations, see
Eq.~(\ref{my_fluid_equations}), separately for electrons and each ion species.
These equations are complemented by Poisson's Eq.~(\ref{Poisson_general}) for
the electrostatic potential.

Rigorously speaking, the 5-moment fluid model given by Eq.
(\ref{my_fluid_equations}) is invalid beyond the long-wavelength limit (LWL)
defined by Eq.~(\ref{long_conditions}) and discussed at length in
Sec.~\ref{Long-wavelength limit}, since otherwise the kinetic effects of
Landau damping [not included in Eq.~(\ref{my_fluid_equations})] start playing a
crucial role by suppressing the instability within a sufficiently
short-wavelength range. Nonetheless, exploring the general dispersion relation
given by Eq.~(\ref{General_dispersion_compact}) for arbitrary wavelengths,
even with no regard for kinetic effects, still makes sense because the
fluid-model description is generally much more popular than is a more rigorous
kinetic one. Most importantly, fluid-model simulations require much less
computer resources than do kinetic simulations and they can cover much larger
spatial scales. This would allow one to use global fluid-model codes developed
for large-scale processes for analyzing the small-scale plasma instabilities
as well.

Bearing in mind such possibilities, it is imperative to study the instability
driving conditions within the entire domain of $\vec{k}$, including the limit
opposite to the LWL. The short-wavelength limit has been explored in
Appendix~\ref{APPENDIX A} with an important conclusion that sufficiently short-wavelength
waves are always stable, regardless of how strong is the driving electric
field. It is especially important that this short-wavelength wave stabilization takes
place even in spite of the fact that the fluid equations lack Landau damping.
The unavoidable consequence of the short-wavelength stabilization is the fact
that somewhere between the long-wavelength limit with positive $\gamma\propto
k^{2}$ and the linearly stable short-wavelength limit with $\gamma<0$ there
necessarily exists an absolute maximum of the instability growth rate
(although the position of this maximum in the $\vec{k}$-space may differ
significantly from that determined by a more accurate kinetic analysis).

The general dispersion relation for the multi-fluid plasma with arbitrarily
magnetized ions, see Eq.~(\ref{disperga_snova}) or
(\ref{General_dispersion_compact}), describes the entire span of wavevectors,
but the major thrust of this paper is on the long-wavelength limit (LWL)
explored in Sec.~\ref{Long-wavelength limit}. In addition to the fact that
this is the only limit fully justified for the fluid-model approach, this
limit also provides the minimum threshold field for all instabilities. Note
that the threshold value for the ETI requires even stronger wavelength
restrictions given by Eq.~(\ref{SWL}). The LWL
also allows one to separate different instability driving and damping
mechanisms as separate linear contributions to the total growth/damping rate,
see Eqs.~(\ref{gamma_struct}) and (\ref{gamma_DL,IT,ET}). This makes the
physical analysis of the instability drivers much easier.

The major result of any linear theory is the instability threshold because
only if the instability driver exceeds the minimum threshold value
then the instability develops. We present the 5-momentum multi-fluid model
calculations of the instability threshold field in
Sec.~\ref{Threshold electric field}, along with the simpler particular cases.
When the minimum instability threshold is exceeded and hence the instability
develops, the largest values of the growth rate indicate which modes are, at
least initially, the fastest growing. The corresponding wavevectors usually
depend on how well above the threshold is the driving field. In the framework
of our model, however, the position of the fastest growing mode may be
physically inaccurate because we have not included the kinetic effect of
Landau damping. This is especially important for the FBI (and, to some degree,
for the ITI) driving because the ETI driving is automatically maximized at a
loose interface between the superlong-wavelength limit (SLWL) and the LWL,
i.e., assuredly within the field of applicability of the fluid model.

Using a fully kinetic PIC code, recently we simulated collisional
instabilities for the solar chromosphere parameters and, to our total
surprise, found that ETI may be a dominant instability in the solar
chromosphere \cite{Oppenheim:Newly20}, The paper by Gogoberidze et al.
\cite{Gogoberidze:Electrostatic14} has also stressed the importance of the ETI
in solar chromosphere, albeit from a somewhat different perspective (unlike
Ref.~\onlinecite{Gogoberidze:Electrostatic14}, we have not included Coulomb
collisions in this model). According to our analysis, one can safely
assume that the purely multi-fluid description of the ETI, unlike the FBI,
driving is reasonably accurate.

Results of these studies can be used for simple predictions of collisional
instabilities in various low-ionized plasma media, like the solar
chromosphere. One of the most important findings is the statement that the
5-moment fluid-model equations will necessarily provide damping of
sufficiently short-wavelength waves, regardless of the driving field strength.
This allows one to safely employ global fluid codes developed for modeling
large-scale processes to model small-scale collisional plasma instabilities,
even though the kinetic effect of Landau damping is not included. Using the multi-fluid
code Ebysus~\cite{Martinez-Sykora:On20}, we have
already started such modeling for the solar chromosphere\cite{Evans:Multi02_arxiv}.
Reference~\onlinecite{Evans:Multi02_arxiv} also includes
comparison with the analytic theory.

\appendix
\renewcommand{\thesection}{\Alph{section}}
\renewcommand{\thesubsection}{\fnsymbol{subsection}}
\section{SHORT-WAVELENGTH LIMIT\label{APPENDIX A}}

This appendix discusses the short-wavelength limit of the general dispersion
relation. This analysis is important because its results assure that the
employed fluid model, even without Landau damping, can be safely used for
instabillity modeling with no need for additional damping mechanisms to
stabilize the wave behavior at short wavelengths.

We define the short-wavelength limit (SWL) by assuming%
\end{subequations}
\begin{equation}
\omega,kV_{Tj},kV_{0},\omega_{Ds}\gg\nu_{sn}\gtrsim\delta_{sn}\nu_{sn},
\label{SWL_conditions}%
\end{equation}
while, for simplicity, the wavelength remains still much longer than the Debye
lengths, $k\lambda_{Ds}\ll1$. Under conditions of Eq.~(\ref{SWL_conditions}),
using $\delta_{sn}\lesssim1$, we have%
\begin{equation}
\frac{1}{\mu_{e}}\approx1-i\ \frac{\delta_{en}\nu_{en}}{\omega_{De}}%
,\qquad\frac{1}{\mu_{j}}\approx1-i\ \frac{\delta_{jn}\nu_{jn}}{\omega_{Dj}}.
\label{1/mu_SWL}%
\end{equation}
Since $\delta_{en}\ll1$, in what follows we will neglect the electron cooling,
$1/\mu_{e}\approx1$, but will retain the ion cooling with the energy loss
fraction, $\delta_{jn}=2m_{j}/(m_{j}+m_{n})$, typically of order unity. In
what follows, we neglect the thermal instability drivers described by $B_{s}$
since thermal perturbations easily spread out over the short-wavelength waves
due to the heat advection, even within the LWL, as we discussed in
Sec.~\ref{Long-wavelength limit}, and hence will not be destabilizing within
the SWL. The heat conductivity, not included in
Eq.~(\ref{my_fluid_equations_temperat}), will even increase this temperature
spread. This leaves us with the only instability driver, namely, the FBI one.

For small $\left\vert \nu_{sn}/\omega_{Ds}\right\vert $, in accord with the
conditions imposed by Eq.~(\ref{SWL_conditions}), we obtain%
\begin{widetext}
\[
A_{e}\approx-\ \frac{T_{e}k^{2}V_{Tj}^{2}\psi_{j}}{T_{j}\nu_{jn}\nu_{en}%
}\left(  1+\frac{i\nu_{en}}{\omega_{De}}\right)  ,\qquad A_{j}\approx
\frac{k^{2}V_{Tj}^{2}}{\omega_{Dj}^{2}}\left(  1-\frac{i\nu_{jn}}{\omega_{Dj}%
}\right)  ,
\]
so that%
\begin{eqnarray*}
 \frac{\rho_{j}\alpha_{j}A_{j}}{A_{e}} &\approx& -\ \frac{\rho_{j}\nu_{jn}%
\nu_{en}}{\psi_{j}\omega_{Dj}^{2}}\left[  1-i\left(  \frac{\nu_{en}}%
{\Omega_{e}}+\frac{\nu_{jn}}{\omega_{Dj}}\right)  \right]  ,\\
 \frac{1-\left[  1+2/\left(  3\mu_{e}\right)  \right]  A_{e}}{1-\left[
1+2/\left(  3\mu_{j}\right)  \right]  A_{j}}
  &\approx& \frac{1+5T_{e}/(3T_{j})\left(  1+i\nu_{en}/\omega_{De}\right)
k^{2}V_{Tj}^{2}\psi_{j}/(\nu_{jn}\nu_{en})}{1-(5/3)k^{2}V_{Tj}^{2}\left[
1-i\left(  1+2\delta_{jn}/5\right)  \nu_{jn}/\omega_{Dj}\right]  /\Omega
_{j}^{2}}.
\end{eqnarray*}

As a result, Eq.~(\ref{General_dispersion_compact}) becomes%
\begin{equation}
  D(\omega_{Dj}^{2})=1-\sum_{j=1}^{p}\frac{\rho_{j}}{\omega_{Dj}^{2}}\left[
1-i\left(  \frac{\nu_{en}}{\omega_{De}}+\frac{\nu_{jn}}{\omega_{Dj}}\right)
\right]\frac{\nu_{jn}\nu_{en}/\psi_{j}+\left[  5T_{e}/(3T_{j})\right]
\left(  1+i\nu_{en}/\omega_{De}\right)  k^{2}V_{Tj}^{2}}{\omega_{Dj}%
^{2}-(5/3)k^{2}V_{Tj}^{2}\left[  1-i\left(  1+2\delta_{jn}/5\right)  \nu
_{jn}/\omega_{Dj}\right]  }=0. \label{D_promo}%
\end{equation}
Assuming, in addition to conditions (\ref{SWL_conditions}),%
\begin{equation}
k^{2}V_{Tj}^{2}\gg\frac{3\nu_{jn}\nu_{en}}{5\psi_{j}}\ \frac{T_{j}}{T_{e}}%
\geq\frac{3\Omega_{j}\Omega_{e}}{5}\ \frac{T_{j}}{T_{e}}, \label{additionally}%
\end{equation}
in the long numerator of Eq.~(\ref{D_promo}) we neglect the term $\nu_{jn}%
\nu_{en}/\psi_{j}$. Then, keeping the same linear accuracy with respect to
$\nu_{sn}/\omega_{Ds}$ as above, we reduce Eq.~(\ref{D_promo}) to a simpler
relation:%
\begin{equation}
D(\omega_{Dj}^{2})=1-\sum_{j=1}^{p}\frac{5\rho_{j}T_{e}}{3T_{j}}%
\ \frac{\left[  1-i\left(  \nu_{jn}/\omega_{Dj}\right)  \right]  k^{2}%
V_{Tj}^{2}}{\Omega_{j}^{2}-(5/3)k^{2}V_{Tj}^{2}\left[  1-i\left(
1+2\delta_{jn}/5\right)  \nu_{jn}/\omega_{Dj}\right]  }=0. \label{D_simpler}%
\end{equation}

\subsection{Phase-velocity relations (the zeroth-order approximation)}

To the zeroth-order approximation, after neglecting all small terms
proportional to $i\nu_{sn}$, the dispersion relation (\ref{D_simpler}) reduces
to%
\begin{equation}
D(\omega_{Dj}^{2})\approx D_{0}(\omega_{Dj}^{2})=1-\sum_{j=1}^{p}\frac
{5\rho_{j}T_{e}}{3T_{j}}\ \frac{k^{2}V_{Tj}^{2}}{\omega_{Dj}^{2}%
-(5/3)k^{2}V_{Tj}^{2}}=0. \label{SWL_zero-order_simplified}%
\end{equation}
\end{widetext}
This provides the lowest-order approximation for $\omega_{Dj}$ which also
automatically becomes its dominant real part, $\left(  \omega_{Dj}\right)
_{r}=\operatorname{Re}(\omega_{Dj})$.

For single-species ions (SSI), $j\rightarrow i$, $p=1$, $\rho_{i}=1$, we
obtain the standard phase-velocity expression for ion-acoustic waves,
\begin{equation}
\left(  \omega_{Di}\right)  _{r}=kC_{s},\qquad C_{sA}=\left[  \frac{5}%
{3}\left(  \frac{T_{e}+T_{i}}{m_{i}}\right)  \right]  ^{1/2}, \label{kC_s}%
\end{equation}
where $C_{sA}$ is the ion-acoustic speed for both electrons and ions in the
adiabatic regime (in the isothermal regime, $5/3$ would be replaced by $1$).
Equation~(\ref{kC_s}) can be interpreted as the phase-velocity relation
because it provides the expression for the wave frequency
$\omega=(\omega_{Ds})_{r}+\vec{k}\cdot\vec{V}_{s0}$ and the corresponding wave phase velocity
$(\vec{V}_{\mathrm{ph}})_{i}=\omega/k_{i}$.

Similarly to the zeroth-order equation discussed in Sec.~\ref{Wave phase velocity relation}, in the general case of multi-species ions,
Eq.~(\ref{SWL_zero-order_simplified}) reduces to the $p$-th order polynomial
equation for the unknown quantity $\omega_{Dj}^{2}$ ($p$ is the
total number of ion species). Different values of $V_{Tj}^{2}$ make the
analytical solution of Eq.~(\ref{SWL_zero-order_simplified}) either
complicated (for $p=2,3,4$) or, in general, impossible ($p\geq5$). As will be seen below,
the specific values of $\omega_{Dj}^{2}$ play no role for the main conclusion of this
appendix.

\subsection{Growth/damping rates (the first-order approximation)}

To the next-order accuracy, we include the terms proportional to the small
parameters $i\nu_{sn}/\omega_{Ds}$ as first-order additions. This will give
rise to the small imaginary addition to the wave frequency,
$\omega_{Ds}=(\omega_{Ds})_{r}+i\gamma$, i.e., to the wave
growth/damping rate (since $\gamma$ is the imaginary part of $\omega$ it is
the common imaginary part of all $\omega_{Ds}$). Within the small terms
$\propto i\nu_{sn}/\omega_{Ds}$, we can replace $\omega_{Ds}$ by its dominant
real parts $(\omega_{Ds})_{r}$, though for the sake of brevity
we will keep for the latter the simplest notation, $\omega_{Ds}$. When and
where $\omega_{Ds}$ are the full complex Doppler-shifted wave frequencies or
when they mean their dominant real parts will be clear from the context.

Note that the simple procedure of separation of the dominant real part and the
small imaginary part becomes only possible because in the SWL the absolute
value of the growth/damping rate, $|\gamma|$, automatically turns out to be small
compared to $(\omega_{Ds})_{r}$. This situation is
similar to the opposite long-wavelength limit, $\omega_{Ds}\ll\nu_{sn}$,
formally for the same mathematical reasons, but under different physical
conditions. In the intermediate range of $|\omega_{Ds}|\sim\nu_{sn}$, where
the instability growth rate often reaches its maximum, we should not
necessarily expect $|\gamma|$ to always be much less than $(\omega_{Ds})_{r}$.
Note also that any first-order real corrections to
the zeroth-order values of $\omega_{Ds}$ will be of no interest to us because
they would lead only to small corrections in the wave phase-velocity relation
without affecting in any appreciable way the growth/damping rates.

Now we return to the full reduced dispersion relation (\ref{D_simpler}).
Linearizing it by including the remaining small terms $\propto i\nu
_{sn}/\omega_{Ds}$, as well as $i\gamma$ within the dominant real parts of the
equation, we can rewrite this equation as%
\begin{widetext}
\begin{equation}
\sum_{j=1}^{p}F_{j}\left(\omega_{Dj}\right)  =1,\qquad
F_{j}\left(  \omega_{Dj}\right)  = \frac{5\rho_{j}T_{e}}{T_{j}}
\ \frac{\left[  1-i\left(  \nu_{jn}/\omega_{Dj}\right)\right]
k^{2}V_{Tj}^{2}}{3\omega_{Dj}^{2}-5k^{2}V_{Tj}^{2}\left[1-i\left(  1+2\delta
_{jn}/5\right)  \nu_{jn}/\omega_{Dj}\right]}. \label{via_F}%
\end{equation}
\end{widetext}
To the first-order accuracy with respect to the small parameters $i\nu
_{sn}/\omega_{Ds}$ and $i\gamma/\omega_{Ds}$, expanding each
$F_{j}(\omega_{Dj}t)$ in Taylor series to the first-order (linear) terms, we
obtain%
\[
F_{j}\left(  \omega_{Dj}\right)  \approx F_{j0}(\omega_{Dj})+i\gamma\left.
\frac{\partial F_{j0}}{\partial\omega_{Dj}}\right\vert _{\omega_{Dj}=(
\omega_{Dj})_{r}}+i\operatorname{Im}F_{j}(\omega_{Dj})
,
\]
where $F_{j0}$ is the function $F_{j}(\omega_{Dj})  $ with
neglected terms $\propto i\nu_{sn}/\omega_{Ds}$, $F_{j0}(\omega
_{Dj})  \approx\operatorname{Re}F_{j}((\omega_{Dj})_{r})$,
while the argument of $i\operatorname{Im}F_{j}(\omega_{Dj})$ still includes full
$\omega_{Dj}$ with linear $i\nu_{sn}/\omega_{Ds}$ corrections. Assuming that we know all
roots $\omega_{Dj}\approx(\omega_{Dj})_{r}$ of the zeroth-order
equation $\sum_{j=1}^{p}\operatorname{Re}F_{j}(\omega_{Dj})=1$, for each of these $n$ roots we have the equation
\[
i\gamma\sum_{j=1}^{p}\frac{\partial F_{j0}}{\partial\omega_{Dj}}+i\sum
_{j=1}^{p}\operatorname{Im}F_{j}\left(  \omega_{Dj}\right)  =0,
\]
yielding%
\begin{equation}
\gamma=-\ \left.  \frac{\sum_{j=1}^{p}\operatorname{Im}F_{j}(\omega
_{Dj})}{\sum_{j=1}^{p}\partial\operatorname{Re}F_{j}/\partial\omega
_{Dj}}\right|  _{\omega_{Dj}=(\omega_{Dj})_{r}},
\label{gamma_gen}%
\end{equation}
where $\operatorname{Im}F_{j}(\omega_{Dj})$ with $\omega
_{Dj}=(\omega_{Dj})_{r}$ contain only small linear terms
$\propto\nu_{sn}/\omega_{Ds}$.

According to Eq.~(\ref{via_F}), we have%
\begin{equation}
\operatorname{Re}F_{j}=F_{j0}(\omega_{Dj})\approx\frac{5\rho_{j}T_{e}%
k^{2}V_{Tj}^{2}}{T_{j}\left(  3\omega_{Dj}^{2}-5k^{2}V_{Tj}^{2}\right)  },
\label{F_j0}%
\end{equation}
yielding%
\begin{equation}
\frac{\partial F_{j0}}{\partial\omega_{Dj}}\approx-\ \frac{30\rho_{j}%
T_{e}k^{2}V_{Tj}^{2}\omega_{Dj}}{T_{j}\left(  3\omega_{Dj}^{2}-5k^{2}%
V_{Tj}^{2}\right)  ^{2}}=-\ \frac{6\omega_{Dj}F_{j0}\left(  \omega
_{Dj}\right)  }{3\omega_{Dj}^{2}-5k^{2}V_{Tj}^{2}}. \label{dF_j0/dOmega_j}%
\end{equation}
Expanding the expression for $F_{j}\left(\omega_{Dj}\right)$ in Taylor series
to the linear term $\propto i\nu_{jn}/\omega_{Dj}$, we obtain
\begin{equation}
\operatorname{Im}F_{j}\left(  \omega_{Dj}\right)  \approx-\ \frac{i\nu_{j}%
}{\omega_{Dj}}\ \frac{3\omega_{Dj}^{2}+2k^{2}V_{Tj}^{2}\delta_{jn}}%
{3\omega_{Dj}^{2}-5k^{2}V_{Tj}^{2}}\ F_{j0}\left(  \omega_{Dj}\right)  ,
\label{imagi}%
\end{equation}
so that Eqs.~(\ref{gamma_gen})-(\ref{imagi}) yield%
\begin{align}
\gamma &  \approx-\ \frac{\sum_{j=1}^{p}\frac{\nu_{jn}}{\omega_{Dj}}%
\frac{3\omega_{Dj}^{2}+2k^{2}V_{Tj}^{2}\delta_{jn}}{3\omega_{Dj}^{2}%
-5k^{2}V_{Tj}^{2}}\ F_{j0}\left(  \omega_{Dj}\right)  }{2\sum_{j=1}^{p}%
\frac{3\omega_{Dj}}{3\omega_{Dj}^{2}-5k^{2}V_{Tj}^{2}}\ F_{j0}(\omega_{Dj})  }\nonumber\\
&  =-\ \frac{\sum_{j=1}^{p}\frac{\nu_{jn}}{\omega_{Dj}}\ \frac{5\rho_{j}%
T_{e}k^{2}V_{Tj}^{2}\left(  3\omega_{Dj}^{2}+2k^{2}V_{Tj}^{2}\delta
_{jn}\right)  }{T_{j}\left(  3\omega_{Dj}^{2}-5k^{2}V_{Tj}^{2}\right)  ^{2}}%
}{2\sum_{j=1}^{p}\frac{15\rho_{j}T_{e}k^{2}V_{Tj}^{2}\omega_{Dj}}{T_{j}\left(
3\omega_{Dj}^{2}-5k^{2}V_{Tj}^{2}\right)  ^{2}}}. \label{gamma_fin}%
\end{align}
In particular, in the SSI case ($p=1$, $j\rightarrow i$), we have%
\begin{equation}
\gamma\approx-\ \frac{\nu_{i}\left(  3\omega_{Di}^{2}+2k^{2}V_{Ti}^{2}%
\delta_{in}\right)  }{6\omega_{Di}^{2}}. \label{gamma_SSI}%
\end{equation}
These expressions clearly demonstrate that in the SWL the growth/damping rate
$\gamma$ is always negative, regardless of the driving electric field
amplitude. This means that in the large-$k$ limit all waves are absolutely
stable. Hence, somewhere in the
intermediate range between the LWL and SWL, there must be some optimal values
of $\vec{k}$ where the instability growth rate reaches one or several maxima
and then goes down to the negative values described by Eqs.~(\ref{gamma_fin})
or (\ref{gamma_SSI}). This leads to the conclusion that the employed fluid
model can be safely used for instabillity modeling with no need for any
additional damping mechanisms at short wavelengths to stabilize there wave
behavior. Though this analysis has neglected a few minor factors, such as the
charge separation, etc., the neglected factors are mostly wave-stabilizing and
could not change the main conclusion.

\section*{Appendix B: LIST OF MAJOR NOTATIONS\label{APPENDIX B}}

\noindent$A_{s}$, is defined by Eq.~(\ref{ABmu}), see also
Eq.~(\ref{A_s_opiat});

\noindent$B_{s}$, is defined by Eq.~(\ref{ABmu}), see also Eq.~(\ref{B_s_gen});

\noindent$\vec{B}_{0}$ is the external magnetic field ($B_{0}=|\vec{B}_{0}|$);

\noindent$\hat{b}=$ $\vec{B}_{0}/B_{0}$ is the unit vector along $\vec{B}_{0}$;

\noindent$C_{s}$ is the isothermal ion-acoustic speed [see
Eq.~(\ref{V_0_approx})];

\noindent$D(\omega,\vec{k})$ is the dispersion function in the LWL, see
Eq.~(\ref{D_reduced})\ [$D_{0}(\omega_{r},\vec{k})$ is the dominant real part
of $D(\omega,\vec{k})$, Eq.~(\ref{D_0})];

\noindent$\vec{E}_{0}$ is the external electric and magnetic field
($E_{0}=|\vec{E}_{0}|$);

\noindent$E_{\mathrm{Thr}}$ is the instability threshold field;

\noindent$E_{\mathrm{Thr}}^{\min}$ is the temperature-modified minimum FBI
threshold field [see Eq.~(\ref{7})];

\noindent$N_{s}$ is defined by Eq.~(\ref{phi=snova});

\noindent$F(\zeta_{e})$, see Eq.~(\ref{F,mu_a});

\noindent$G_{j}$ is the quantity defined in Eq.~(\ref{kU_via_G});

\noindent$\vec{K}_{s}=\delta\vec{V}_{s}/\left(  \alpha_{s}\phi+\eta_{s}%
+\tau_{s}\right)  $ is a temporary notation used in
Sec.~\ref{Linear wave perturbations: General dispersion relation};

\noindent$\vec{k}$ is the wavevector ($k=|\vec{k}|$\ is the wavenumber);

\noindent$M_{sn}=m_{s}m_{n}/(m_{s}+m_{n})$ is the effective mass of the two
colliding particles ($s$ and $n)$;

\noindent$m_{s}$ is the $s$-species particle mass;

\noindent$n_{s}$ is the $s$-species particle number density;

\noindent$p$ is the total number of the ion species;

\noindent$q_{s}$ is the $s$-species particle electric charge ($q_{e}=-e$);

\noindent$R$ is defined by Eq.~(\ref{RR});

\noindent$T_{s}$ is the $s$-species particle temperature (in energy units);

\noindent$\vec{U}_{j}\equiv\vec{V}_{e0}-\vec{V}_{j0}$ is the difference
between the undisturbed electron and ion drifts [see Eq.~(\ref{UU})];

\noindent$\vec{V}_{0}$ is the $\vec{E}_{0}\times\vec{B}_{0}$-drift velocity;

\noindent$\vec{V}_{s0}$ is the $s$-species mean fluid velocity;

\noindent$V_{Ts}=(T_{s0}/m_{s})^{1/2}$ is the mean thermal speed of the
$s$-species particles;

\noindent$\vec{V}_{\mathrm{ph}}=\omega/\vec{k}$ is the wave phase velocity

\noindent$\alpha_{s}\equiv T_{e0}q_{s}/(T_{s0}e)$ is a temporary parameter
introduced in Eq.~(\ref{alpha_s});

\noindent$\alpha_{n}$ is the neutral-particle polarizability, Eq.~(\ref{MMC});

\noindent$\gamma$ is the wave growth/damping rate;

\noindent$\delta A\propto\exp[i(\vec{k}\cdot\vec{r}-\omega t)]$ denotes a
harmonic wave perturbation of any scalar or vector quantity $A$ ($A_{0}$ is
the undisturbed value);

\noindent$\delta_{sn}$ is the mean collisional energy-loss fraction
($\delta_{sn}=\delta_{sn}^{\mathrm{elas}}=2m_{s}/(m_{s}+m_{n})$ for elastic collisions);

\noindent$\epsilon_{0}$ is the permittivity of free space;

\noindent$\varepsilon$ is a small parameter, see Eq.~(\ref{epsilon});

\noindent$\zeta_{s}=\Omega_{s}/kV_{0}$ is a normalized quantity introduced in
Sec.~\ref{Wave phase velocity relation}\ (there $\Omega_{s}\approx\Omega_{sr}$);

\noindent$\eta_{s}$ is a normalized perturbation of the $s$-species particle
density, $n_{s}$ [see Eq.~(\ref{Introducing})];

\noindent$\Theta_{j}=(\kappa_{j}/\kappa_{e})^{1/2}$ is a small parameter
introduced in Sec.~(\ref{Theta_j});

\noindent$\theta$ is the angle (in radians) from $\vec{V}_{0}$ to $\vec{k}$
(the `flow' angle);

\noindent$\kappa_{s}=\omega_{cs}/\nu_{sn}$ is the magnetization ratio of the
$s$-species particles;

\noindent$\lambda_{Ds}=[\epsilon_{0}T_{s0}/(e^{2}n_{s0})]^{1/2}$ is the
`partial' Debye length of the $s$-species;

\noindent$\mu_{s}$ is a complex quantity introduced in Eq.~(\ref{ABmu});

\noindent$\nu_{sn}$ is the mean collision frequency of the $s$-species
particles with neutrals;

\noindent$\xi_{j}$, see Eq.~(\ref{F,mu_a});

\noindent$\rho_{j}=(q_{j}/e)(n_{j0}/n_{e0})$ is introduced in
Eq.~(\ref{Debye_snova});

\noindent$\sigma_{sn}$ is the $s$-$n$ collisional cross-section;

\noindent$\tau_{s}$ is a normalized perturbation of the $s$-species particle
temperature, $T_{s}$ [see Eq.~(\ref{Introducing})];

\noindent$\Phi$ is the electrostatic potential;

\noindent$\phi$ is a normalized perturbation of the electrostatic potential
$\Phi$ [see Eq.~(\ref{Introducing})];

\noindent$\chi_{j}=\theta+\arctan\kappa_{j}$ is an angle (in radians), see
also Eq.~(\ref{ion_angles});

\noindent$\psi_{j}$ is the quantity defined by Eq.~(\ref{psi_j});

\noindent$\Psi$ is the quantity defined by Eq.~(\ref{Psi});

\noindent$\omega_{Ds}\equiv\omega-\vec{k}\cdot\vec{V}_{s0}$ is the
Doppler-shifted frequency in the frame of reference moving with the
$s$-species mean flow, $\vec{V}_{s0}$ [see Eq.~(\ref{Omega_alpha})];

\noindent$\Omega_{s}$ is the gyrofrequency of the $s$-species particles;

\noindent$\omega=\omega_{r}+i\gamma$ is the wave frequency (both $\omega_{r}$
and $\gamma$ are real);

\noindent Subscripts $\parallel$ and $\perp$ relate to the vector components
parallel and perpendicular to $\vec{B}_{0}$, respectively.

\begin{acknowledgments}
We acknowledge the support of this work by NSF Grant No. 1903416, NASA grants
80NSSC20K1272, 80NSSC21K0737, 80NSSC21K1684, and contract NNG09FA40C.
\end{acknowledgments}

\section*{DATA AVAILABILITY}

Data sharing is not applicable to this article as no new data were created or
analyzed in this study.
%
%

\end{document}